\documentclass[fleqn,usenatbib]{mnras}
\usepackage{gensymb}
\usepackage{comment}
\usepackage{newtxtext,newtxmath}
\usepackage[T1]{fontenc}

\usepackage[utf8]{inputenc}
\usepackage{graphicx}	
\usepackage{amsmath}	
\usepackage{relsize}
\usepackage{booktabs}
\usepackage{xcolor}
\usepackage{subcaption}
\DeclareRobustCommand{\VAN}[3]{#2}
\let\VANthebibliography\thebibliography
\def\thebibliography{\DeclareRobustCommand{\VAN}[3]{##3}\VANthebibliography}

\newcommand{\omegai}{\mbox{$\omega_\mathrm{i}$}}
\newcommand{\PM}{\mbox{$\mathrm{\mathbf{PM}}$}}
\newcommand{\fbin}{\mbox{$f_\mathrm{bin}$}}
\newcommand{\logtyr}{\mbox{$\log(t/\mathrm{yr})$}}
\newcommand{\Msun}{\mbox{$\mathrm{M}_\odot$}}
\newcommand{\sigmaPLR}{\mbox{$\sigma_{-\ln(\text{PLR})}$}}

\title[]{Strikingly high fraction of fast rotators in Magellanic Cloud star clusters}

\author[]{Greta Ettorre$^{1,2,3}$,
Alessandro Mazzi$^{1,2}$,
Léo Girardi$^{4}$,
Paola Marigo$^{1,}$\thanks{Passed away on October 20, 2024},
Giada Pastorelli$^{4}$, 
Paul Goudfrooij$^{5}$,\newauthor
Benjamin F. Williams$^{6}$,
Andrea Bellini$^{5}$, 
Alessandro Bressan$^{7}$, 
Yang Chen$^{8,9}$,
Matteo Correnti$^{10,11}$,\newauthor
Guglielmo Costa$^{1,12,4}$,
Julianne J. Dalcanton$^{13,6}$, 
Pietro Facchini$^{14}$,
Morgan Fouesneau$^{15}$, \newauthor
Chi Thanh Nguyen$^{16}$, 
Guglielmo Volpato$^{1}$
\\
$^{1}$Dipartimento di Fisica e Astronomia "Galileo Galilei", Università di Padova, Vicolo dell’Osservatorio 3, I-35122 Padova, Italy\\
$^{2}$Dipartimento di Fisica e Astronomia “Augusto Righi”, Alma Mater Studiorum, Università di Bologna, Via Gobetti 93/2, I-40129 Bologna, Italy\\
$^{3}$INAF - Osservatorio di Astrofisica e Scienza dello Spazio di Bologna, Via Gobetti 93/3, 40129 Bologna, Italy\\
$^{4}$INAF - Osservatorio Astronomico di Padova, Vicolo dell’Osservatorio 5, I-35122 Padova, Italy\\
$^{5}$Space Telescope Science Institute, 3700 San Martin drive, Baltimore, MD, 21218, USA\\
$^{6}$Department of Astronomy, University of Washington, Box 351580, Seattle, WA 98195, USA\\
$^{7}$SISSA, International School for Advanced Studies, Trieste, Italy\\
$^{8}$School of Physics and optoelectronic engineering, Anhui University, Hefei 230601, China\\
$^{9}$National Astronomical Observatories, Chinese Academy of Sciences, Beijing 100101, China\\
$^{10}$INAF - Osservatorio Astronomico di Roma, Via Frascati 33, 00078, Monteporzio Catone, Rome, Italy\\
$^{11}$ASI - Space Science Data Center, Via del Politecnico, I-00133, Rome, Italy\\
$^{12}$Université de Lyon, Univ Lyon1, ENS de Lyon, CNRS, Centre de Recherche Astrophysique de Lyon UMR5574, F-69230 Saint-Genis-Laval, France\\
$^{13}$Center for Computational Astrophysics, Flatiron Institute, 162 Fifth Avenue, New York, NY 10010, USA\\
$^{14}$Astronomisches Rechen-Institut, Zentrum f\"ur Astronomie der Universit\"at Heidelberg, M\"onchhofstr. 12-14, D-69120 Heidelberg, Germany\\
$^{15}$Max Plank Institute for Astronomy (MPIA), K{\"o}nigstuhl 17, 69117 Heidelberg, Germany\\
$^{16}$INAF - Osservatorio Astronomico di Trieste, via Tiepolo 11, 34143 Trieste, Italy\\
}

\date{Accepted XXX. Received YYY; in original form ZZZ}

\pubyear{2025}

\begin{document}
\label{firstpage}
\pagerange{\pageref{firstpage}--\pageref{lastpage}}
\maketitle

\begin{abstract}
There has been growing evidence that the rich star clusters in the Magellanic Clouds contain significant fractions of rapidly rotating stars. In this work, we aim to constrain these fractions by studying the colour-magnitude diagrams of four star clusters, selected among those with the most striking signatures of fast rotators. Using isochrones derived from \textsc{parsec v2.0} stellar tracks, we generate distinct stellar populations, each covering a limited interval of initial rotation rates $\omegai$, referred to as `Partial Models' (PMs). Using optimization algorithms and Monte Carlo Markov Chains, PMs are combined to create the final best-fitting model. In our analysis, we adopt two key assumptions: a uniform age and an isotropic distribution of stellar spin axes within each cluster. The solutions are allowed to explore the entire range of $\omegai$, and different values of age, metallicity, distance and foreground extinction. We find that the rotational velocity distributions in all four clusters reveal a high fraction of stars with $\omegai$ close to the break-up value, in all cases. Specifically, the fraction of stars with $\omegai>0.7$ exceeds 80\% in the clusters NGC 419 of the Small Magellanic Cloud (SMC) and NGC 1831 and NGC 1866 of the Large Magellanic Cloud (LMC). For NGC 2203 of the LMC, this fraction is smaller, although it still exceeds $50\%$, confirming that also this cluster is mainly populated by fast-rotating stars.
\end{abstract}


\section{Introduction}\label{introduction}

In the last years, the high-quality photometry obtained with the  
\textit{Hubble Space Telescope} (\textit{HST}) revealed the presence of peculiar features in the colour-magnitude diagrams (CMDs) of young and intermediate-age clusters in the Magellanic Clouds (MCs). These include extended main-sequence turn-offs (eMSTOs), split main sequences (MSs) and dual red clumps (RCs).

One of the first hypotheses in this regard was the so-called ``age-spread scenario'', according to which the eMSTO is due to stars in the parent cluster that have ages spread from 100 to 500 Myr \citep{Mackey2008,Milone_2009,Goudfrooij2014,Correnti_2014,Goudfrooij_2015}.  For instance, \cite{Goudfrooij2014} studied a sample of 18 intermediate-age clusters, finding that the distribution of stars across the eMSTO is influenced by two main factors: the star formation histories of the clusters and the mass loss that they experience after the explosion of massive stars in the central regions. Within this framework, they identified a relationship between the width of the MSTO and the escape velocity at the cluster centre, which is crucial for the cluster's capacity to retain or accumulate gas and stellar yields. These results, along with a significant correlation found by \citet{Goudfrooij2014} between the fraction of stars in the bluest part of the MSTO and those in the faint extension of the RC, provide observational evidence supporting the age spread scenario.

On the other hand, the ``stellar rotation scenario'', originally suggested by \citet{Bastian}, explains the eMSTO as the consequence of a spread in rotation velocities among turn-off stars of a coeval population (see  \citealt{GirardiMiglio2011,Georgy_2014,2015Brandt,2015Niederhofer,Georgy2019} for a more detailed discussion).
Indeed, the presence of rotating stars in MCs clusters has been confirmed by several spectroscopic studies performed with the VLT \citep{Marino2018a,Kamann2018,kamann20,kamann23}  
and, for instance, \cite{Dupree2017} found direct evidence of a population of rapidly rotating stars in NGC 1866 from the spectra of 29 MSTO stars obtained at the Magellan-Clay telescope.
Moreover, recent studies performed with \textit{Gaia} \citep{Marino2018b,Cordoni2018} have revealed the presence of the eMSTO feature also in Milky Way (MW) open clusters. In this framework, \cite{Cordoni2024} studied 32 galactic open clusters finding that the eMSTO morphology is accurately reproduced by a single population with a high rotation rate, observed with rotation axis inclinations ranging from 0$\degree$ to 90$\degree$.

Rotation can have a remarkable impact on stellar structure and evolution, inducing geometrical distortion and extra mixing \citep{Maeder2009}. All these effects have been widely described in the literature, starting from the early work by \citet{VonZeipel1924}.
Following the \cite{Maeder1997} scheme to treat rotation in stellar codes, the departure from spherical symmetry induced by rotation is proportional to the rotation rate $\omega=\Omega/\Omega_{\mathrm{c}}$, where $\Omega$ is the angular velocity and $\Omega_{\mathrm{c}}=(2/3)^{3/2}\sqrt{GM/(R_{\mathrm{pol}})^3}$, with $\textit{G}$ the gravitational constant and $R_{\mathrm{pol}}$ the polar value of the stellar radius, is the critical break-up value, that is the angular velocity at which the centrifugal force balances the gravity at the equator of a rotating star of mass $\textit{M}$. The distortion is a result of the centrifugal force lowering the local gravity of the star, resulting in an effective gravity, $g_{\mathrm{eff}}$, that varies along the surface. Therefore, $g_{\mathrm{eff}}$ becomes dependent on the rotation rate $\omega$ and the colatitude $\theta$. 
Following the \citeauthor{VonZeipel1924} theorem, the local effective temperature of the star, $T_{\mathrm{eff}}$, is proportional to the effective gravity, hence it inherits the dependence on $\omega$ and $\theta$. This effect is known as ``gravity darkening effect'' and introduces a new variable when computing the outcoming flux of the star, which is the inclination angle $i$ of the star rotation axis with respect to the observer's line of sight. A star deformed by fast rotation appears brighter and hotter if viewed pole-on ($i=0\degree$) rather than equator-on ($i=90\degree$). The fundamental equations are presented in \citet{Girardi_2019_rotation}. 
Other than geometrical distortion, rotation produces instabilities that redistribute the chemical elements and angular momentum throughout the star, causing extra mixing of the stellar material. All these combined effects can alter both the colour and magnitude of stars. As a result, the CMD of a cluster containing rotating stars is expected to exhibit distinctive features related to rotation.

\begin{figure*}
    \centering
    \includegraphics[width=\textwidth]{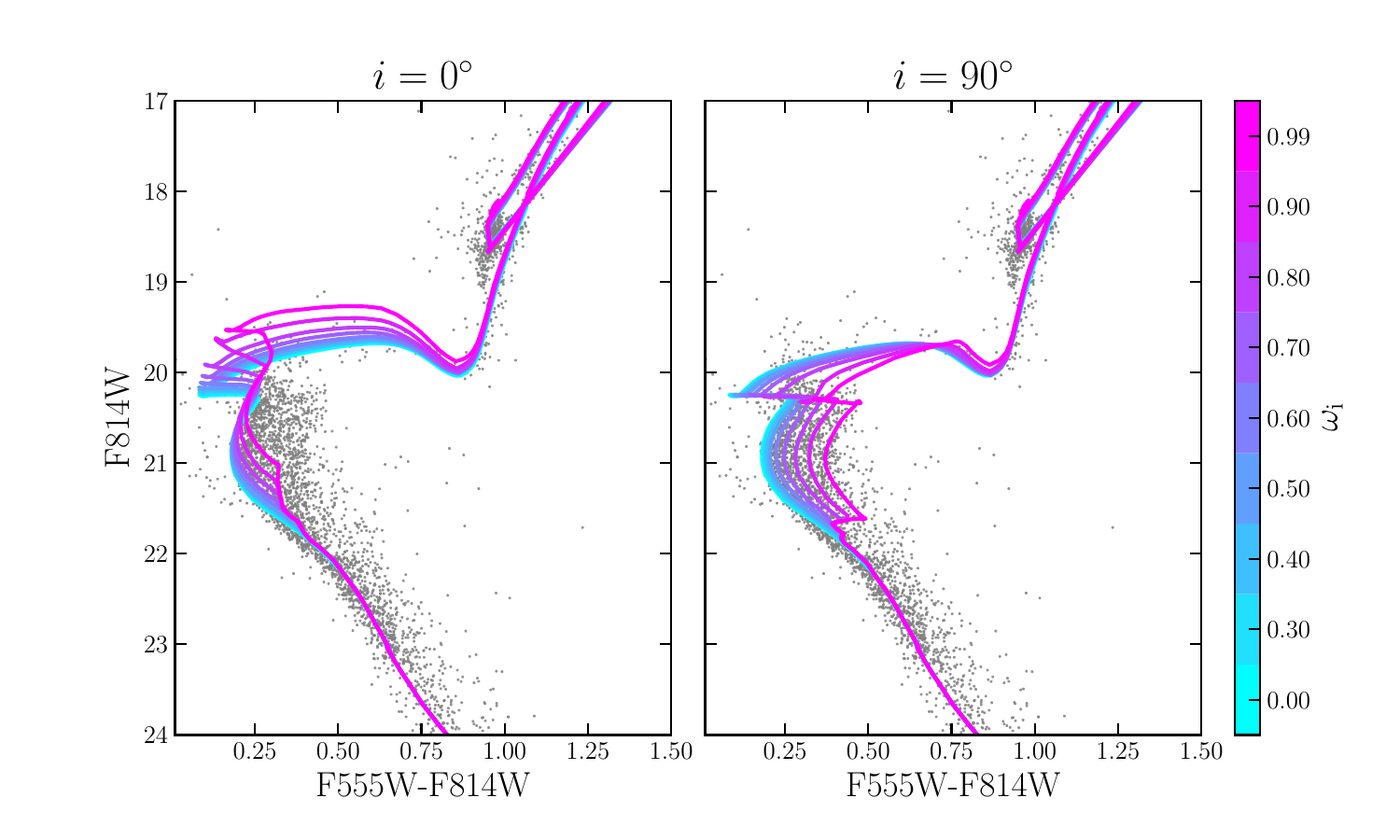}
    \caption{Left: the CMD for NGC 419 as derived from the Advanced Camera for Surveys (ACS) data centered on the cluster (see Sect.~\ref{419} for the details). The overlaid isochrones correspond to a metallicity of $Z=0.004$, age $\logtyr=9.125$, inclination $i=0 \degree$, true distance modulus $(m-M)_0=18.89$, absorption in $V$ band $A_{\mathrm{v}}=0.15$ and initial rotation rate from $\omegai =0.0,0.3,0.4,0.5,0.6,0.7,0.8,0.9,0.99$. Right: the same as the left-hand panel but with inclination $i=90 \degree$.}
    \label{fig9}
\end{figure*}

In Fig.~\ref{fig9}, we illustrate how rotation affects the colour and magnitude by superimposing isochrones corresponding to different initial rotation rates, $\omegai$, on top of a stellar cluster's CMD, focusing on the eMSTO. This is done for the two extreme values of the inclination angle, $i=0 \degree$ and $i=90 \degree$. 
The isochrones are produced with \textsc{trilegal} \citep{girardi05TRILEGAL} starting from \textsc{parsec v2.0} \citep{PARSEC} evolutionary tracks.
At the MSTO, when $i=0 \degree$ (pole-on configuration), isochrones with a high rotation rate are brighter than isochrones with lower $\omegai$, as expected from both the mixing processes and the deformation induced by rotation. 
On the other side, when $i=90 \degree$, isochrones corresponding to fast rotation are much redder than isochrones with lower $\omegai$.
Moreover, it is worth noticing that, when $i=90 \degree$, the evident spread of the MSTO could be easily reproduced with isochrones corresponding to stellar populations with the same age but different initial rotation rate $\omegai$.

\begin{figure}
    \centering
    \includegraphics[width=\hsize]{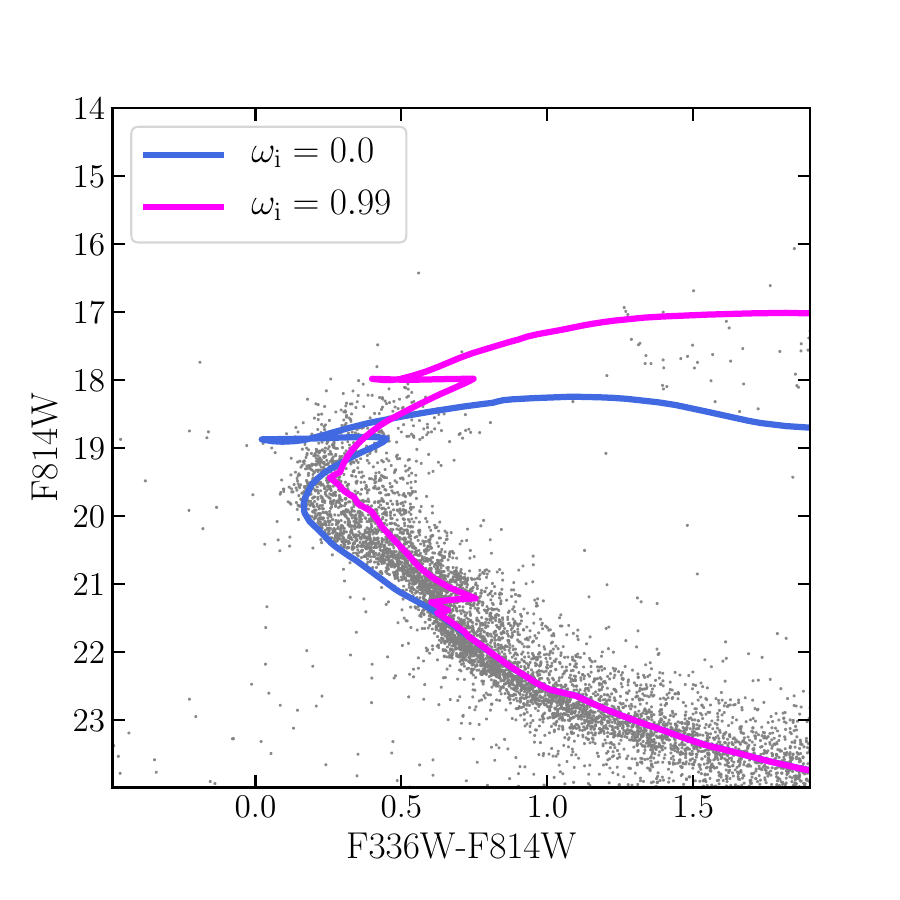}
    \caption{The colour-magnitude diagram of NGC 1831 as derived from \textit{HST} WFC3 observation with overplotted \textsc{parsec v2.0} isochrones for two values of the initial rotation rate $\omegai=0.0$ (in blue) and $\omegai=0.99$ (in magenta). Above F814W $\sim 21$ there is a sudden change of the MS slope, which becomes slightly bluer and broader. This feature is called MS kink and is attributed to the onset of strong convection in the outer layers of stars with masses below those at the kink \citep{Goudfrooij2018}. This causes their rotation to slow-down through magnetic braking. 
    Therefore, the kink would represent an empirical measure of the stellar mass above which rotation sets in.
    The MS kink is incorporated in \textsc{parsec v2.0} isochrones, as clearly visible in the magenta one.
    The value of the inclination angle for the rotating isochrone is $i=90\degree$}
    \label{mskink}
\end{figure}

Another aspect to be considered is that fast rotation moves the location of the MS as a whole, but only above a certain brightness. Indeed, as pointed out by \citet{Goudfrooij2018}, some MC clusters present a kink at the MS, above which the MS becomes slightly bluer and appears slightly broader. This is illustrated in Fig.~\ref{mskink}, where \textsc{parsec v2.0} isochrones are plotted on top of NGC 1831 data (see Sect.~\ref{ngc1831} for details) for the two extreme values of the initial rotation rate $\omegai=0.0$ (in blue) and $\omegai=0.99$ (in magenta), and a value of the inclination angle of $i=90\degree$ in both cases. The feature of the MS kink is clearly visible in the magenta isochrone at F814W $\sim21$.
\citet{Dantona2002} noticed that the full spectrum of turbulence convection model produces a sudden change of the MS slope at the onset of strong convection in the outer layers of stars, which turns out to coincide with the level of the MS kink in NGC 1831, suggesting that the kink is related to this process \citep{Goudfrooij2018}. The kink occurs at a stellar mass for which the convective envelope suddenly deepens much more into the interior than that of a star with a mass only $0.01\,\Msun$ larger, thus causing a significant decrease of the temperature dependence of stellar mass, $\text{d}T_{\mathrm{eff}}/\text{d}M$, with decreasing stellar mass. 
\citet{Goudfrooij2018} suggest that the onset of strong convection in the outer layers of the lowest-mass and faintest stars causes their rotation to slow-down through magnetic braking. 
Therefore, the kink would represent an empirical measure of the stellar mass above which rotation sets in. According to \citet{Goudfrooij2018}, in NGC 1831, this occurs at a mass of $1.45 \pm 0.02\,\Msun$ and F814W $\sim21$. The feature can be inserted into isochrones only in an approximate way. For instance, \textsc{parsec v2.0} models do it by assuming a gradual growth of rotation along an initial mass interval of $\Delta M_\mathrm{i}=0.3\,\Msun$ \citep[see sect. 2.4 in][]{PARSEC}.

Due to the presence of these striking CMD features, MC clusters present fantastic opportunities to test and calibrate stellar models with rotation and to probe the environments in which fast-rotating stars arise.
CMDs of MC clusters have been quantitatively interpreted with the use of rotating stellar models starting with \citet{Bastian}. Initial works were affected by the incomplete modelling of the problem; for instance, \citet{Bastian} used evolutionary tracks instead of isochrones in their analysis, hence ignoring the change in stellar lifetimes caused by rotation, whereas \citet{GirardiMiglio2011} used isochrones but did not model the strong spread in colours caused by viewing extremely-fast-rotating stars at different inclinations. The CMD modelling of the eMSTO phenomenon started to become more reliable when more extended grids of rotating models were made available by the SYCLIST \citep{Georgy_2014,Georgy2019} and MIST \citep{MIST,Gossage2018} teams, as for instance those used in \citet{BrandtHuang} and \citet{Gossage2019}. 

Notable examples of CMD analyses using stellar population models that incorporate fast-rotating stars include the following studies. \citet{Gossage2018} employed the MESA stellar evolution code to create models with rotation rates spanning from $\omega = 0.0$ to 0.6, to examine how stellar rotation impacts the inferred ages of open clusters, specifically focusing on the Hyades, Praesepe, and Pleiades clusters. 
\citet{Gossage2019} extend the stellar models up to $\omega = 0.9$ and fit the MSTO regions of five Magellanic Cloud clusters finding that a rotation-rate distribution alone qualitatively matches the observed eMSTO structures. In another study, \citet{Lipatov2022} calculated continuous probability densities in three-dimensional colour, magnitude, and $v\sin i$ space for individual stars in the eMSTO of NGC 1846, to jointly infer age dispersion, rotational distribution, and binary fraction of the cluster, based on MIST rotating stellar evolution models. Additionally, \citet{Wang2022} used MESA models to explore the origin of the blue MS in four young clusters of the MCs, proposing that blue MS stars primarily form through stellar mergers. Lastly, \citet{Wang2023} evaluated the amount of rotation needed to account for the observed colour split in the split MS of young star clusters, using newly developed MESA models, along with the SYCLIST and MIST models. 
The results obtained by \citet{Lipatov2022} suggest that their models should be modified reducing the magnetic braking for low-mass stars. Meanwhile, \citet{Wang2022} found that stars on the blue MS appear bluer than predicted by even their non-rotating isochrones. Furthermore, \citet{Wang2023} reported that stellar rotation velocities inferred from isochrone comparisons were highly model-dependent, arguing that moderately fast rotation be sufficient to explain the observed colour spreads in MC clusters. Hence, all these works find some level of discrepancy between models and observations, indicating that current stellar evolution models may require further refinement to accurately reproduce observed features in CMDs.
In fact, there is still ample room for improving the modeling, because many aspects, such as the onset of magnetic braking, the efficiency of rotational mixing, and the computation of the emerging flux in fast-rotating stars, are still either being calibrated or are subject to well-known simplifications. Moreover, previous works have focused on the CMD region around the eMSTOs, neglecting the constraints provided by red giants, and often not discussing the quantitative constraints provided by the lower MS. 

In this context, the extension of the quantitative analysis using additional sets of models can only be beneficial.
In this work, we investigate the distribution of stellar rotation rates within four MC clusters with a new set of models derived from \textsc{parsec v2.0} \citep{Costa_thesis_2019,Costa2019,PARSEC} evolutionary tracks. 
The paper is organized as follows: in Sect.~\ref{data}, we introduce the star clusters studied in this work. In Sect.~\ref{methods}, we describe the main tools used in this analysis and the steps that were required to get the results, presented in Sect.~\ref{results}. Finally, in Sect.~\ref{conclusion}, we draw our conclusions.

\section{Target clusters}\label{data}

For this work, we choose four densely populated clusters with well-defined features in their CMDs, which are currently interpreted as signatures of the presence of fast rotators: NGC 419 in the Small Magellanic Cloud (SMC) and NGC 2203, NGC 1831, and NGC 1866 in the Large Magellanic Cloud (LMC).
Table~\ref{tab:paramsMC} contains the primary parameters initially adopted for each cluster,
encompassing age, metallicity, true distance modulus, and foreground extinction. The same table presents the origin of our data, the softwares used for the photometric reduction and the central coordinates and radius used to select the subsamples we use in our analysis.

\begin{figure*}
    \centering
    \begin{minipage}{0.49\textwidth}
    \includegraphics[width=\columnwidth]{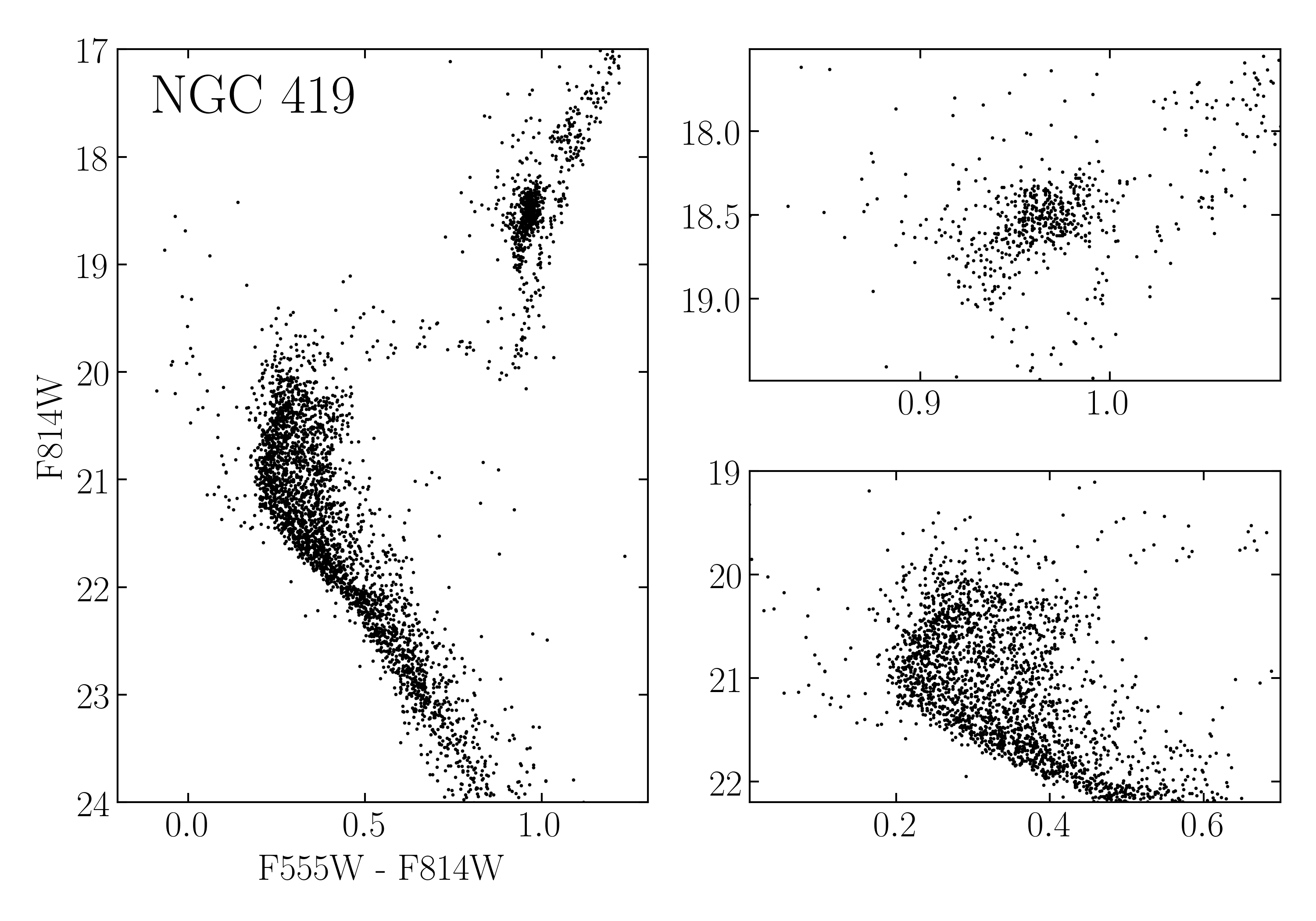}
    \includegraphics[width=\columnwidth]{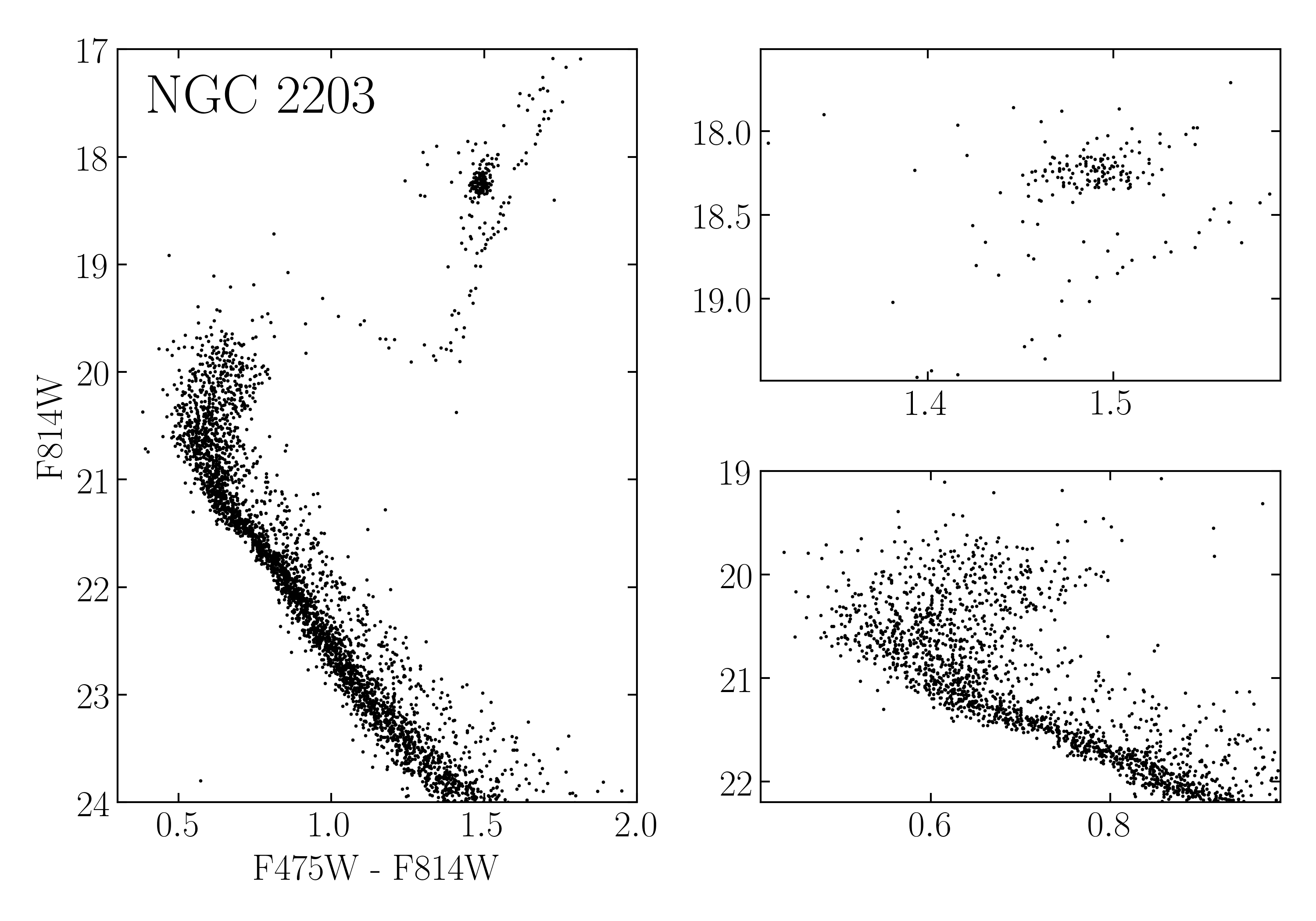}
    \end{minipage}
    \begin{minipage}{0.49\textwidth}
    \includegraphics[width=\columnwidth]{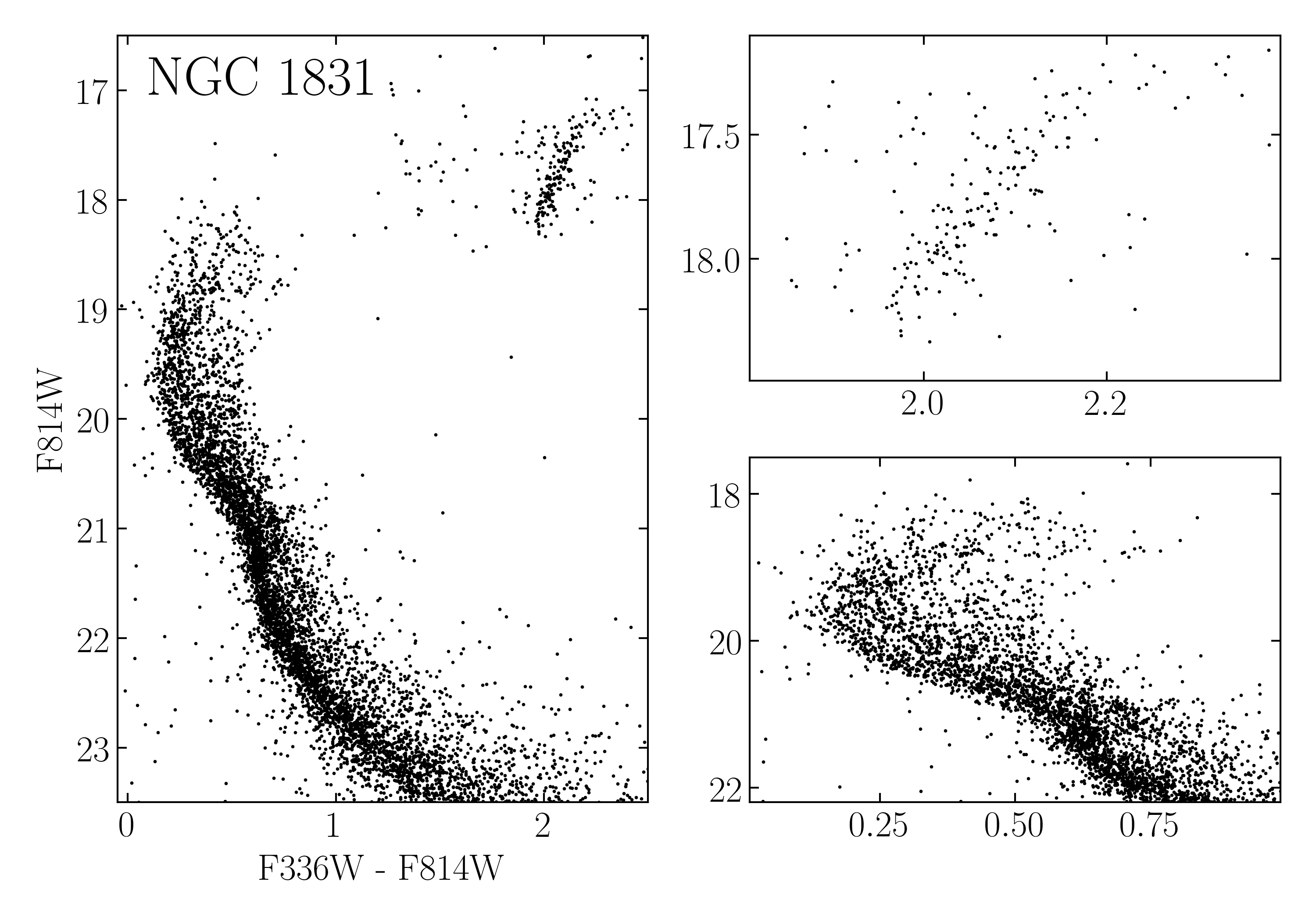}
    \includegraphics[width=\columnwidth]{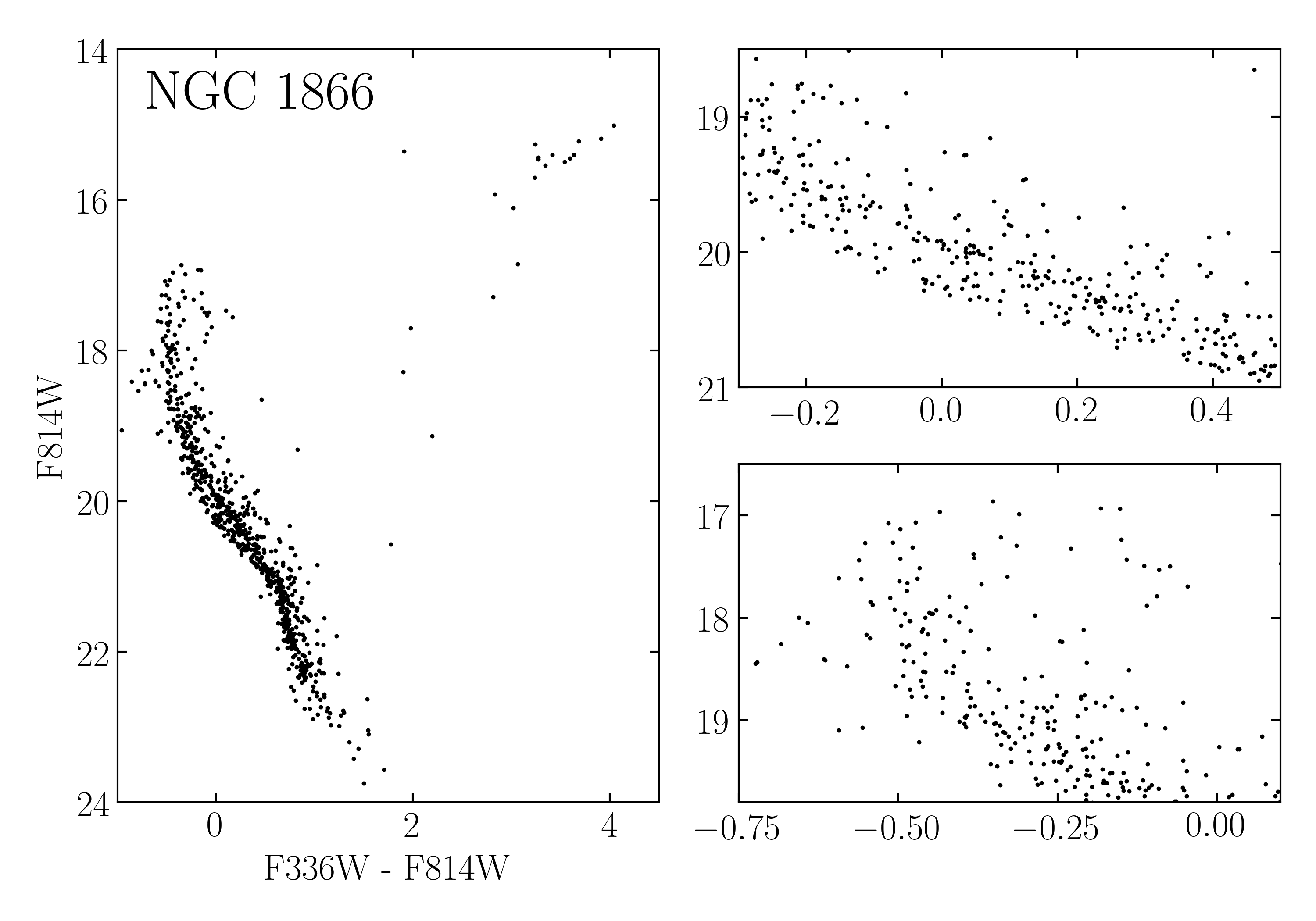}
    \end{minipage}
    \caption{The CMDs of the four selected MC clusters. In all cases, the left panel gives an overview of the CMD, while the right panels zoom at the features of most interest for our analyses. They include the MSTO region and the RC, except for NGC 1866, where we zoom on the MSTO and on the bottom MS.}
    \label{obsCMDs}
\end{figure*}

Fig.~\ref{obsCMDs} illustrates the complete CMDs selected for our analysis, zooming onto the CMD features that are currently interpreted as the clear signatures of the presence of fast-rotating stars.

\begin{table*}
\caption{Main parameters and origin of data of our target clusters.}
\label{tab:paramsMC}
\begin{tabular}{|l|c|c|c|c|}
\toprule
Parameter & NGC 419 & NGC 2203 & NGC 1831 & NGC 1866\\ 
\midrule
Age (Gyr) & $1.45 \pm 0.05$ $^1$ & $1.55 \pm 0.05$ $^1$ & $\sim 0.8$ $^2$ & $\sim 0.25$ $^3$ \\ 
$[\mathrm{Fe/H}]$ (dex) & $-0.7 \pm 0.1$  $^1$ & $-0.3 \pm 0.1$  $^1$  & $-0.25$ $^3$ & $-0.36$ $^4$ \\ 
$(m-M)_0$ (mag)  & $19.1$ $^9$ & $18.37 \pm 0.03$  $^1$ & $18.35$ $^2$ & $18.43$ $^3$ \\ 
$A_{\mathrm{v}}$ (mag) & $0.15 \pm 0.02$  $^1$ & $0.16 \pm 0.02$  $^1$ & $0.11$ $^2$ & $0.28$ $^3$ \\ 
R.A.$_\mathrm{c}$ & $01^\mathrm{h}08^\mathrm{m}17.2^\mathrm{s}$ $^5$ & $06^h 04^m 42.0^s$ $^5$ & 05 06 17.4$^7$ & 05 13 38.9 $^7$\\
Dec.$_\mathrm{c}$ & $-72\degree53'01''$ $^5$ & $-75\degree26'18''$ $^5$ &  -64 55 11 $^7$ & -65 27.  52 $^7$ \\
radius (arcsec) & 20 & 40 & 35.2 $^8$ & 29.3 $^8$\\
GO program & 10396 & 12257 & 14688 & 14204,14069\\
PI & Gallagher & Girardi & Goudfrooij & Milone,Bastian\\
camera & ACS/WFC & WFC3/UVIS & WFC3/UVIS & WFC3/UVIS,ACS/WFC \\
filters & F555W,F814W & F475W,F814W & F336W,F814W & F336W(WFC3),F814W(ACS)\\
photometry & Dolphot$^6$ & Dolphot$^6$ & ePSF$^3$ & ePSF$^3$ \\
\bottomrule
\end{tabular}
\\References and comments: $^1$ \citet{Goudfrooij2014}; $^2$
\citet{Correnti2021}; $^3$ \citet{Goudfrooij2018}; $^4$\citet{Gossage2019}; $^5$ From SIMBAD database (REF), ICRS (J2000); $^6$ as in \citet{williams14} and \citet{rosenfield17}; $^7$ \citet{mclaughlin05}; $^8$ that is one effective radius; $^9$we initially adopted $(m-M)_0=18.85$\,mag as in \citet{Goudfrooij2014}, but soon realised that it was too short. This might be due to the fact that the recalibration included a correction for imperfect charge transfer efficiency (CTE), which was not used by \citet{Goudfrooij2014}. Additionally, \citet{Goudfrooij2014} used the \citet{Marigo08} isochrones, which have a different metallicity scale compared to \textsc{parsec} isochrones. \end{table*}

\subsection{NGC 419}\label{419}
NGC 419 is an intermediate-age star cluster ($t=1.45 \pm 0.05$ Gyr; \citealt{Goudfrooij2014}) located to the east of the SMC bar in a region relatively devoid of dust. The NGC 419 data were processed with DOLPHOT \citep{dolphot} using the pipeline by \citet{williams14}. A large number of artificial-star tests (ASTs) were performed as described in Sect.~\ref{pms_spread}. The final data were then cut with the "good star" (GST) criteria defined by \citet{williams14}, with the addition of a cut in a crowding-related parameter (namely CROWD\,$<0.2$) to eliminate the spurious objects detected on the PSF wings of very bright stars. For the final analysis, we select only the stars within 20 arcsec of the cluster centre. 
The final photometric catalog of NGC 419 (Fig.~\ref{obsCMDs}, upper left panel) is clearly dominated by cluster members, but on the other hand, it suffers from severe incompleteness at faint magnitudes due to crowding. The 50 per cent completeness level is located at F814W $\sim22$~mag. Very prominent features are the eMSTO \citep{Glatt2008} and the presence of a well-populated secondary RC \citep{girardi2009}. 

\subsection{NGC 2203}
NGC 2203 is an intermediate-age open cluster ($t=1.55 \pm 0.05$ Gyr, as derived by \citealt{Goudfrooij2014}) and is the oldest object among our targets. 
The data for NGC 2203 were processed in the same way as 
NGC 419.
For the final sample, we select only the stars within 40 arcsec from the cluster centre. 
The cluster's CMD (Fig.~\ref{obsCMDs}, bottom left panel) shows a clear eMSTO, but unlike NGC 419, the RC appears as a single feature without a prominent secondary RC.

\subsection{NGC 1831}\label{ngc1831}
NGC 1831 is a $\sim 800$ Myr old open cluster \citep{Correnti2021} of the LMC.
The photometric analysis was performed as in \citet{Goudfrooij2018}, where the stellar photometry measurements were performed using point-spread function (PSF) fitting with the "effective PSF" (ePSF) package for WFC3/UVIS, based on the ePSF package for the ACS/WFC camera as described by \citet{AndersonKing2006}. Moreover, to minimize the contamination from LMC field stars, we included stars within the effective radius from the cluster center. 
Its CMD, plotted in Fig.~\ref{obsCMDs}, upper right panel, shows the presence of an eMSTO and a quite extended RC -- as expected for a cluster whose stars have avoided electron degeneracy prior to the core-helium burning phase \citep{Girardi2016}.
Another prominent feature in the F336W and F814W data of NGC 1831 is a kink along the MS at F814W $\sim21$\,mag. This feature is much less evident in CMDs involving filters redder than F336W.

\subsection{NGC 1866}

NGC 1866, with an age of approximately $250$ Myr \citep{Goudfrooij2018}, is the youngest cluster of this study, and it is located at the northern periphery of the LMC disc. Similar to NGC 1831, the photometric analysis of this cluster was performed as described in \citet{Goudfrooij2018}. 
Due to its position, its CMD remains relatively unaffected by interstellar dust and field stars ({Fig.~\ref{obsCMDs}, bottom right panel}).
As described in \citet{milone17}, in addition to an eMSTO, NGC 1866 presents a split MS, where a feeble MS appears at bluer colours than the main one. The blue MS is not as evident in our CMD due to the fact that we selected stars within $29.3$ arcsec from the cluster center. In fact, as pointed out by \citet{milone17}, the number of blue MS stars significantly increases going from the central regions of the cluster outwards, reaching up to $\sim 45 \%$. Their figure 8 highlights the changes in the CMD in radial bins, with the blue MS appearing gradually as the distance from the centre increases.
According to \cite{Dantona2017}, the bluer MS can only be interpreted as being the classical MS caused by slowly rotating stars. This leads to the conclusion that the stars belonging to the most prominent MS in NGC 1866 -- and hence most of the stars in NGC 1866, except for the least-massive ones -- are fast rotators. This conclusion is also consistent with the masses inferred for the numerous Cepheids in this cluster (see \citealt{Costa2019Cepheids}).

\section{Methods}\label{methods}
This study aims to analyze the rotational velocity distribution of stars in the four MC star clusters described in Sect.~\ref{data}. We apply a method that can be generally described as "CMD fitting" or "CMD reconstruction", in which the observed CMD is compared to a model CMD made by a linear combination of its sub-components. More specifically, throughout this work, we deal with CMDs in the form of Hess diagrams, which display the stellar density across the CMD.

In our case, the model $\textbf{M}$ is such that:
\begin{equation}\label{model}
    \textbf{M} = \mathbf{PM}_{0} +\sum_j a_j \, \mathbf{PM}_{j}   \,\,\,.
\end{equation}
where $\mathbf{PM_0}$ is a model for the background field of the LMC or SMC and the MW foreground, as detailed in Sect.~\ref{PM0}.  The $\mathbf{PM}_j$ are ``partial models'' (PMs) corresponding to stellar populations covering a limited interval of initial rotation rates $\omegai$ (Sections~\ref{parsec} to~\ref{pms_bin}).  Finally, the coefficients $a_j$ are the weights to be assigned to every $\mathbf{PM}_j$. They are proportional to the amount of stars in the rotational velocity bin $j$, hence they constitute the distribution of $\omegai$ that we want to determine via CMD fitting.

As will be detailed below, three main assumptions are adopted: 
\begin{enumerate}
\item Cluster stars are assumed to be coeval and initially chemically-homogeneous. Therefore, all $\mathbf{PM}_j$ in a model are computed for the same age and metallicity. Importantly, as we treat these two parameters as unknowns, we take care that a suitable interval of age and metallicity is explored for each cluster.
\item The initial mass function and fraction of binaries is the same among all cluster stars (hence the same for all $\mathbf{PM}_j$). 
\item Rotating stars have an isotropic distribution of rotation axes. 
\end{enumerate}
Concerning the second point, assuming a constant binary fraction may be at odds with certain scenarios proposed to explain different rotation rates, such as tidal interactions in close binaries \citep{Dantona2015} or stellar mergers \citep{Wang2022}. However, observational studies of young clusters, such as the one conducted by \cite{Kamann2021} on NGC 1850, appear to support a constant binary fraction.

\subsection{Isochrones for rotating stars}\label{parsec}

We adopt the stellar evolutionary tracks from \textsc{parsec}, the PAdova and tRieste Stellar Evolutionary Code \citep{BRESSAN2012}. 
The effect of rotation was introduced in \textsc{parsec} by \cite{Costa2019}, leading to the version \textsc{2.0} of the code and to the extended library of evolutionary tracks described in \citet{PARSEC}. This library takes into account masses ranging from $0.09\,\Msun$ to $14\,\Msun$ and metallicities between $Z = 0.004$ and $Z = 0.017$, for seven different initial rotation rates in the range $\omegai = 0.00-0.99$. The tracks start on the pre-main sequence, and end at the beginning of the thermally pulsing asymptotic giant branch (TP-AGB) phase for low- and intermediate-mass stars, or at the C-exhaustion for higher masses. For the present work, we use an extended version of the \citet{PARSEC} tracks, covering the entire metallicity range between $Z = 0.002$ and $0.03$ (corresponding to $\mathrm{[Fe/H]}$ values from $-0.89$ to $+0.34$ dex).

Starting from the sets of evolutionary tracks, isochrones are then constructed with the \textsc{trilegal} code \citep{girardi05TRILEGAL}, which has been adapted to interpolate all the additional quantities needed to characterise rotating stars. 
Interpolations are performed not only in the initial mass, age and initial metallicity space (as usual in the calculation of \textsc{parsec} isochrones), but also in initial rotation rate, \omegai, in the range $0.0 \leq \omegai \leq 0.99$, with the same method used in \cite{Bertelli1994} and \cite{Girardi2000}. This interpolation in \omegai\ represents an improvement over the isochrones described in \cite{PARSEC}.  
The isochrones\footnote{These isochrones are accessible through the web interface \url{http://stev.oapd.inaf.it/cmd}.} contain all the quantities of interest to interpret the observations, including the instantaneous rotation rate $\omega$ and the rotational velocity at the equator.

\subsection{Bolometric corrections}\label{bolomcorrection}
Theoretical quantities derived from stellar evolutionary codes need to be converted into magnitudes across different filters using bolometric corrections (BCs).
For non-rotating stars, tables of BCs are used as a function of effective temperature, surface gravity, and surface chemical composition \citep[see][]{Girardi2002, Chen2019}.
Bolometric correction tables for rotating stars have at least two more parameters to be considered: the rotation rate $\omega$ and the inclination angle $i$ of the line of sight with respect to the stellar rotation axes. These are calculated as in \citet{Girardi_2019_rotation} and implemented in the YBC database of BCs\footnote{\url{https://sec.center/YBC/}} by \citet{Chen2019} . In \textsc{trilegal}, these ``rotating BCs'' are fully adopted for stars with an average surface $T_\mathrm{eff}$ larger than about 5100\,K. For cooler stars, we adopt the BCs computed from non-rotating stars; this is a suitable approximation since their rotation rate slows down to values smaller than $\omega\lesssim0.3$, which ensures negligible departures from spherical symmetry (as shown in fig. 1 of \citealt{Girardi_2019_rotation}). 

This approach ensures that, at least for the hot stars met in the clusters' MS regions of the CMD, we take into full account the change of the flux with $i$ in the several passbands involved, without resorting to BCs computed from a surface-averaged $T_\mathrm{eff}$ at every $i$ (as done, e.g., in \citealt{Georgy_2014} and \citealt{Gossage2018}). 
This is important when we consider filters sampling very different wavelengths and/or wavelengths bluewards of the Balmer jump  \citep{Girardi_2019_rotation}.

As for the filter transmission curves and zeropoints for ACS/WFC and WFC3/UVIS, we follow the recommendations from \citet[][and private communication]{calamida22}. The reference fluxes for Vega magnitudes are taken from the CALSPEC \citep{calspec} file \verb$alpha_lyr_stis_010.fits$.
\subsection{Partial models for single stars}\label{pms}

Before they can be compared to the clusters, isochrones require correction for the true distance modulus $(m-M)_0$ and extinction $A_{\lambda}$.
Given the low values of extinction found in our clusters (Table~\ref{tab:paramsMC}), we can adopt the approximation $A_{\lambda} = A_{\mathrm{v}} \cdot C_{\lambda}$, where $A_{\mathrm{v}}$ is the extinction in the V band and the $C_{\lambda}$ is a coefficient derived from \citet{ODonnell1994} interstellar extinction curve applied to a typical yellow star (the Sun), and made available in the YBC tables \citep{Chen2019}. 

Starting from the isochrones, PMs are generated each encompassing a limited interval of initial rotation rates $\omegai$, with a width of $\Delta\omegai=0.1$.  More specifically, for every age and metallicity, we generate a synthetic population with a total initial mass of $5\times10^5\,\Msun$, following the \citet{kroupa2001} initial mass function, and with a flat distribution of \omegai\ inside the selected $\Delta\omegai=0.1$ interval. These synthetic populations contain information about the stellar fluxes seen at 19 different values of inclination $i$, going from $0\degree$ to $90\degree$ with steps of $5\degree$, corresponding to partial models $\mathrm{\mathbf{PM}}_{j}(i)$. In order to derive a population with an isotropic distribution of orientations, we add the stars as seen from these 19 different values of $i$ into the Hess diagrams, weighting each one as follows 

\begin{equation}
    \mathrm{\mathbf{PM}}_{j} = \sum_{i=0\degree}^{90\degree} \frac{\sin i }{W} \mathrm{\mathbf{PM}}_{j}(i)  \,\,\,,
\end{equation}
where $W = \sum_{i=0\degree}^{90\degree} \sin i $.
This is equivalent to simulating a population with about $10^7\,\Msun$ of stars, each one observed at a different $i$.
This means that higher inclinations have higher probabilities of being observed, peaking at $i=90 \degree$ and decreasing towards $i=0 \degree$, as expected. 
These Hess diagrams are then normalised so that every PM represents a total mass of formed single stars of 1\,\Msun.

\begin{figure*}
    \centering
    \includegraphics[width=\textwidth]{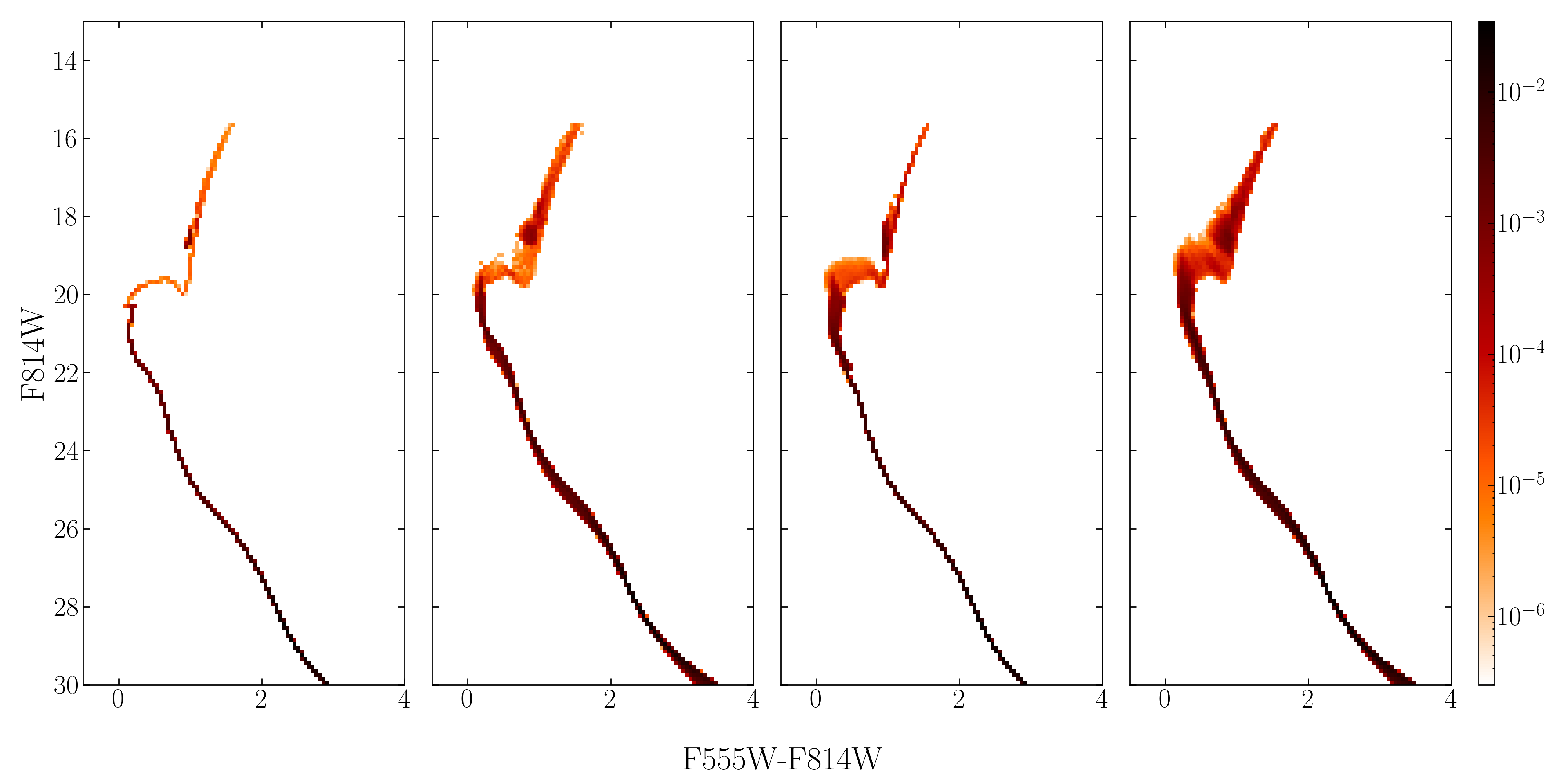}
    \caption{A small subset of the PMs produced for the NGC 419 model at $\logtyr=9.05$ and $Z=0.0035$. They are, from left to right: the PM for $\omegai$ between 0 and 0.1 for single and binary stars, and the PM for $\omegai$ between 0.9 and 0.99 for single and binary stars. The colourbar represents the counts across the Hess diagrams.}
    \label{fig:examplePMs}
\end{figure*}

\subsection{Partial models for binaries}\label{pms_bin}
There are two main fundamental parameters that should be taken into account when generating the partial models for binary stars: the fraction of binaries $f_{\mathrm{bin}}$ and the mass ratio $q = M_2/M_1$.
The former represents the number of binary systems relative to the total number of stars in the cluster, while the latter is the mass ratio of the components, where we assume that $M_2 < M_1$.
When a binary system is not resolved, it appears on the CMD as a single source with the combined light of the two components.
Equal-mass binaries form a sequence nearly parallel to the MS, about $0.75$\,mag brighter.
Unequal-mass binaries appear between the MS of single stars and the line formed by equal-mass binaries, on the red side of the main sequence ridge line (MSRL).

To generate binary stars, a multi-step process is followed. First, the mass of single stars is determined via linear interpolation, and a random number between 0 and 1 is associated to each star. If this number is less than $f_{\mathrm{bin}}$, indicating a potential binary, the binary companion is generated.
The mass ratio (\textit{q}) is then defined using a flat distribution by generating another random number (\textit{n}) between 0 and 1. 
This determines \textit{q} through the equation:
\begin{equation}
    q=q_\mathrm{min}+n \cdot (1-q_\mathrm{min})
\end{equation}
In this work we choose a flat distribution of the mass ratio with $0.7\leq q \leq 1$.
Once the primary star's mass ($M_1$) and $q$ are known, the secondary star's mass ($M_2$) is calculated as $M_2=q \cdot M_1$.
The magnitude of the secondary, $m_2$, is then searched for in the isochrone, assuming it has the same inclination angle as the primary. Finally, the overall binary system magnitude ($m_{\mathrm{bin}}$) is computed as:

\begin{equation}\label{binary}
    m_{\mathrm{bin}}=m_1 - 2.5 \log \Bigl(1+\frac{F_2}{F_1}\Bigr)
\end{equation}
where $F_1=10^{-0.4 m_1}$ and $F_2=10^{-0.4 m_2}$.
As illustrated in Sect.~\ref{sfhfinder}, the fraction of binaries \fbin\ is one of the parameters that are fitted.
For this reason, we generate separate partial models for single and binary stars.
They are subsequently combined with
\begin{equation}
    \mathrm{\mathbf{PM}}_{j} =  (1-\fbin) \,\mathrm{\mathbf{PM}}_{j,\mathrm{single}} +  \fbin \,\mathrm{\mathbf{PM}}_{j,{\mathrm{binary}}} 
\end{equation}

Just for illustrative purposes, Fig.~\ref{fig:examplePMs} presents four of the PMs produced for the study of NGC 419. They cover a much wider range of colours and magnitude than actually necessary to model the cluster data. All evolutionary phases seem to be adequately represented with just minimal Poisson noise. These PMs illustrate the main changes in the MS models as we increase \omegai, and the different mean position of binaries as compared to single stars. Models at this specific age and metallicity are particularly interesting because they illustrate the different RC morphology that can be produced by rotation: the $\omegai<0.1$ case presents a compact RC, while the $\omegai>0.9$ case presents an elongated RC.

\begin{table*}
\centering
\caption{Simulations parameters. From left to right: cluster name, metallicity range and \logtyr\ range in which PMs were computed, name of the VMC subregion used to simulate the field, area of the cluster used in the fit. 
}
\label{tableanalysis}
\begin{tabular}{|c|c|c|c|c|}
\toprule
Cluster name & Metallicity $Z$ range & $\logtyr$ range &VMC tile and subregion(s) & cluster area $(\mathrm{arcmin}^2)$\\
\midrule
NGC 419  & 0.0025 -- 0.0050 & 9.000 -- 9.150 & SMC 4\_4 G6, G10 & 0.349\\

NGC 2203  & 0.0050 -- 0.0090 & 9.100 -- 9.275 & LMC 3\_7 G1 & 1.40\\

NGC 1831  & 0.0050 -- 0.0090 & 8.800 -- 8.975  & LMC 9\_3 G6& 1.08\\

NGC 1866 & 0.0050 -- 0.0090 & 8.400 -- 8.600  & LMC 9\_4 G9 & 0.749\\
\bottomrule
\end{tabular}
\end{table*}

\subsection{Field contribution}\label{PM0}
As shown in Eq.~\ref{model}, the final model $\textbf{M}$ includes $\PM_{0}$, which represents the contribution by the background field of the LMC or SMC and the foreground of the MW.
It is common to estimate such a foreground/background by taking stars observed in a nearby region of the sky with the same area as the cluster region being studied.
But this is often not possible: the areas imaged by the ACS/WFC and WFC3/UVIS cameras are too small to include both a populous MC cluster and a region completely free from its stars. Even if \textit{HST} observations are frequently available in a neighbouring ``parallel field'' located 6 arcmin away from the cluster, the camera and filters of these observations are not the same as in the case of the cluster. Moreover, these nearby areas have crowding conditions very different from the central parts of the clusters, therefore their suitability to represent the cluster foreground/background should not be taken for granted.

To obviate these problems, we produce models for the foreground and background of each cluster with the \textsc{trilegal} code, using the same filters of the observations. In particular, the MW foreground is computed with the calibrations adopted in the \textsc{trilegal} model of the MW \citep{girardi05TRILEGAL, Girardi2016}, which generally makes good predictions for sky areas far from the Galactic Plane \citep[see, e.g., fig. 12 in][]{lauer21}.

The LMC and SMC backgrounds are simulated using the star formation history (SFH) maps that were derived by \citet{Mazzi} and \citet{Rubele2018}, respectively, using data from the VISTA survey of the Magellanic Clouds \citep[VMC;][]{Cioni_2011}.
For the particular case of NGC 2203, which is located outside of the SFH maps, we used the closest VMC field, which is 1.26 deg away from it.
Table~\ref{tableanalysis} contains the names of the simulated regions for each of our target clusters. The map of the VMC tiles and subregions for the LMC is shown in fig. 1 of \cite{Mazzi}.
The Hess diagrams produced from these simulations are then rescaled to match the area of the cluster actually used in the fit.

\subsection{Simulating errors and incompleteness}\label{pms_spread}

After creating partial models for all populations that can be affecting the observed CMD (single stars, binaries, and the cluster foreground/field), we simulate the effects of photometric errors on our models through artificial star tests (ASTs). Even though errors derived with this technique might not reproduce exactly the distribution of the observed ones, in regions of the CMD with high completeness they are still a reasonably good approximation. Therefore, we chose to compute errors by resorting to ASTs: this process involves the generation of artificial stars across the entire CMD, covering the required sky area with a density that follows the observed one. Typically, a few $10^4$ artificial stars are used for every cluster in our sample. These stars are injected into the original \textit{HST} images, which are then reprocessed through the photometric extraction pipeline. From the differences between input and output magnitudes and colours, we derive the position-dependent distribution of photometric errors, while the fraction of non-recovered stars as a function of magnitude gives a measurement of incompleteness. Subsequently, these distributions of errors and incompleteness are applied to the model PMs. 

The final result is a corrected set of PMs, in the form of Hess diagrams, for any given value of age, metallicity and $\omegai$ range, and comprising single stars, binaries, and the cluster foreground/field.

\subsection{Final model}\label{sfhfinder}
The final model $\textbf{M}$ of Eq.~\ref{model} is computed using \textsc{sfhfinder} \citep{Mazzi}. This code was originally conceived to derive spatially-resolved SFH of nearby galaxies, but with minor modifications, it can be applied to our problem as well. 
In order to find the best-fitting parameters for a given dataset the code combines the power of optimization via stochastic gradient descent and the exploration capabilities of MCMC. While the initial gradient descent phase preferentially explores high likelihood regions, the subsequent MCMC phase ensures thorough exploration of the entire parameter space. Moreover, the priors are chosen to be uniform distributions, hence they should not have an impact on our final results.
The first fitted parameters are the $a_j$ coefficients, which tell the relative importance of each partial model $\mathrm{\mathbf{PM}}_{j}$.
The binary fraction \fbin\ is fitted as well. Moreover, we chose to fit both the shift in colour ($\Delta x$) and the shift in magnitude ($\Delta y$), to compute the distance modulus and the extinction of each cluster after the fit. To evaluate the performance of the code, in Appendix~\ref{app:mocks} we show and discuss fits of several mock observations created from the partial models presented above using different input parameters.

\subsection{Simulations and statistical analysis}\label{simulations}
The analysis of each cluster could be performed at fixed values of age and metallicity, after adopting literature values for these parameters. However, we prefer not to do so, because: (1) We will be using new stellar evolutionary models, which have evolutionary times slightly (but significantly) different from those previously used in the literature. In particular, this refers to differences in evolutionary timescales between non-rotating and rotating models. As shown in fig. 3 of \citet{PARSEC}, lifetimes can vary by $\sim 10\%$ for stars with masses under $2\,\Msun$, while they can change by up to 40$\%$ for stars with $M_\mathrm{i}>2\,\Msun$. (2) The metallicities mentioned in the literature are derived either using the fitting of isochrones (and hence have to be recomputed anyway), or using different reference values for the solar metallicity. Therefore, we explore an ample set of values of age and metallicity, compared to those listed in Table~\ref{tab:paramsMC}, with the idea of identifying the best-fitting model, a posteriori, as the one with the maximum likelihood among all cases. 

For each cluster, we conduct the analysis across a number $N_{\mathrm{age}}$ of $\logtyr$ values, evenly spaced at intervals of $\Delta \log t=0.025$, and across $N_Z$ values of metallicity with steps of $\Delta Z=0.0005$.
Therefore, for each cluster we obtain $N_{\mathrm{age}} \cdot N_Z$ best-fitting models $\textbf{M}$ given by equation~\ref{model}.

In all cases, for each couple of age and metallicity 10 partial models are produced, corresponding to 10 intervals of the initial rotation rate: $\omegai$ from 0.0 to 0.1, from 0.1 to 0.2, etc., until the last partial model for \omegai\ in the interval from 0.9 to 0.99.
Table~\ref{tableanalysis} outlines the ranges of metallicity and age within which the analysis is carried out for each cluster, together with the name of the simulated field subregions from the VMC survey, and the area of the cluster that we used in the fit. 

Given the Poissonian nature of our data, the most suitable statistical approach to identify the best model among the $N_{Z} \cdot N_{\mathrm{age}}$ best-fitting models computed for each cluster is to calculate and minimize the Poisson likelihood ratio \citep[see][]{Dolphin2002}:
\begin{equation}\label{PLR}
    - \ln (\text{PLR}) = \sum_k m_k-n_k+n_k \, \ln\frac{n_k}{m_k}
\end{equation}
where $k$ is an index for the bins in the Hess diagram, and $m_k$ and $n_k$ are the star counts in the model and in the data, respectively.

\begin{table*}
\centering
\caption{Properties of the best-fitting models of all clusters. From left to right: cluster name, age and metallicity of the grid point with the highest likelihood, binary fraction, shift in colour $\Delta x$, shift in magnitude $\Delta y$. Distance \textit{d} and V-band extinction $A_\text{V}$ are derived from $\Delta x$ and $\Delta y$.}
\label{fittingparam}
\begin{tabular}{|cc|cccc|cc|}
\toprule
Cluster & \logtyr & $Z$ & \fbin & $\Delta x$ & $\Delta y$ & $d$ & $A_\text{V}$ \\
        & & && (mag) & (mag) & (kpc) & (mag) \\
\midrule
NGC 419  & 9.100 & 0.0035 & $0.140 \pm 0.011$ & $-0.008 \pm 0.001$ & $-0.022 \pm 0.004$ & 65.73 & 0.13\\
NGC 2203 & 9.175 & 0.0075 & $0.104 \pm 0.011$ & $-0.038 \pm 0.001$ & $0.231 \pm 0.005$ & 53.42 & 0.10 \\
NGC 1831 (with kink) & 8.875 & 0.0070 & $0.173 \pm 0.008$ & $-0.019 \pm 0.002$ & $0.165 \pm 0.005$ & 50.71 & 0.09\\
NGC 1831 (no kink) & 8.875 & 0.0070 & $0.173 \pm 0.009$ & $-0.030 \pm 0.003$  & $0.180 \pm 0.006$& 51.14 & 0.08\\
NGC 1866 & 8.500 & 0.0090 & $0.067 \pm 0.018$ & $-0.213 \pm 0.011$ & $0.034 \pm 0.003$ & 52.03 & 0.08 \\
\midrule 
NGC 2203$^1$ & $9.150$ & $0.0060$& $0.075 \pm 0.010$ & $0.105 \pm 0.001$ & $0.286 \pm 0.004$ & $53.58$ & $0.02$ \\
\bottomrule
\end{tabular}
\\$^1$ Alternative solution for NGC 2203 as described in Sect.~\ref{uncertainties}.
\end{table*}

\section{Results and discussion}\label{results}
In this section we discuss the results we obtained for the four clusters of the analysis. In Sect.~\ref{bestfitmodels} we present the likelihood maps and the suggested best-fitting models, in Sect.~\ref{uncertainties} we discuss the uncertainties on age and metallicity of our results and Sect.~\ref{rotatiodistr} summarizes the distribution of rotational velocities we derive.

In Table~\ref{fittingparam} we summarize the properties of the best-fitting models including the age and metallicity of the grid point with the highest likelihood and the parameters obtained from the fit, namely the binary fraction ($\fbin$), the shift in color ($\Delta x$) and the shift in magnitude ($\Delta y$). In addition, the latter two parameters are used to calculate the distance and the V-band extinction of the cluster after the fit, which are included in the last two columns of Table~\ref{fittingparam}.

The results for all clusters are summarized in Figs.~\ref{summary_ngc419}, \ref{summary_ngc2203}, \ref{summary_ngc1831}, \ref{summary_ngc1831_mask} and \ref{summary_ngc1866}.
The top-left panel of each one shows the likelihood map derived from the grid of models, and on top of it we display the minimum of the $- \ln (\text{PLR})$ and the $+ 1 \sigmaPLR$ value of its distribution, computed by finding the interval containing $\sim 68\%$ of the solutions in the corresponding MCMC chain.
The first three levels of the color map correspond to the $1 \times \sigmaPLR$, $2 \times \sigmaPLR$ and $3 \times \sigmaPLR$ values, respectively, while the remaining levels are logarithmically spaced from the third level up to the maximum of the $- \ln (\text{PLR})$. 
Since we are fitting each grid point separately, we do not provide confidence intervals for $\logtyr$ and $Z$ as we do for the fitted parameters listed in Table~\ref{fittingparam}. However, we discuss the uncertainties for both age and metallicity in Sect.~\ref{uncertainties}.

\subsection{Best-fitting models}\label{bestfitmodels}

\begin{figure*}
    \centering
    \includegraphics[width=\textwidth]{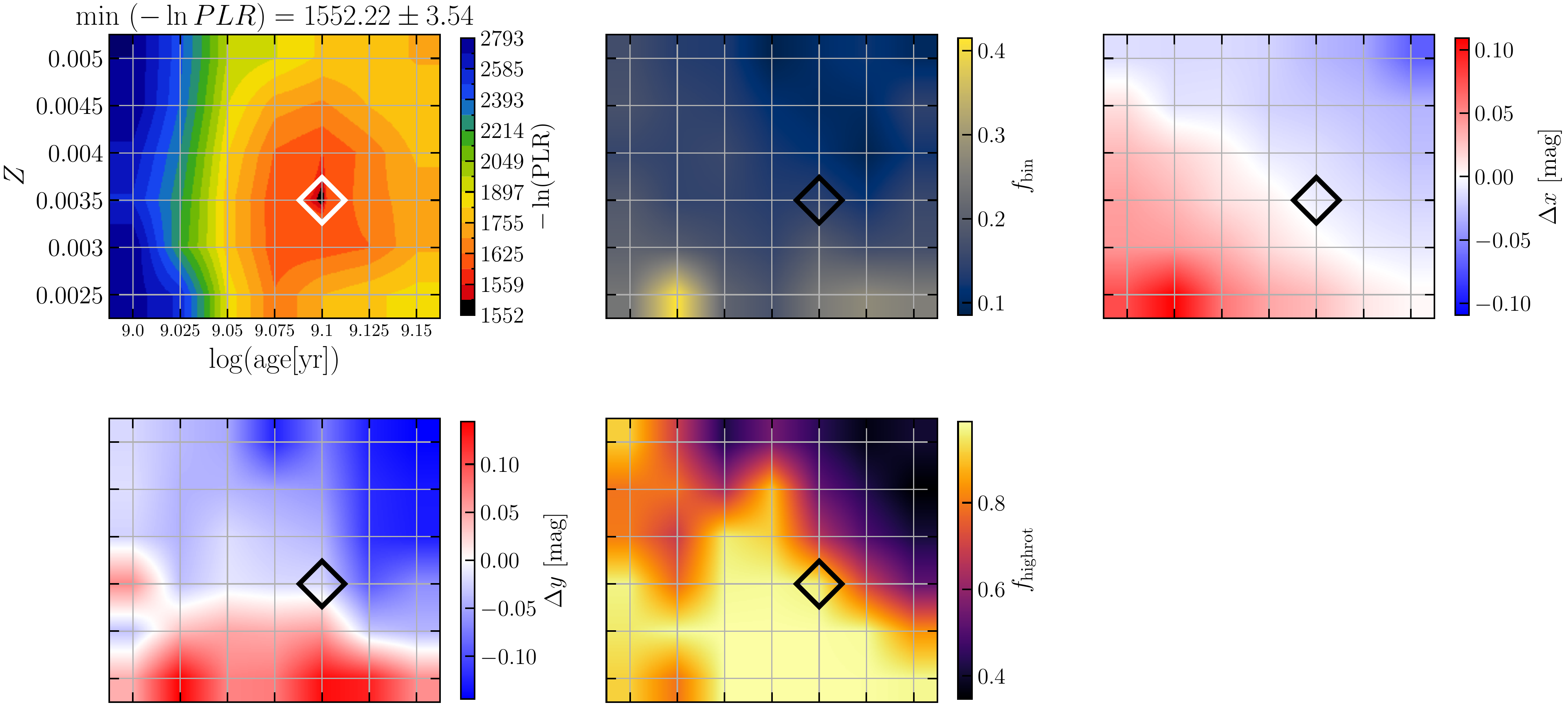}
   \caption{A summary plot showing the results of fitting the CMD of NGC 419, at all age-metallicity grid points, with the parameters introduced in Sect.~\ref{sfhfinder}. Each panel shows a slice through the space of best fits. Moving from left to right, top to bottom, the first panel shows the $- \ln (\text{PLR})$ for the best-fitting models across the grid of log(age) vs. metallicity we explored. More specifically, models were computed at the grid positions, and the plot presents a simple bi-linear interpolation in this grid that is visually represented through a stepped colormap, with levels as detailed in the text. From this panel, it is possible to identify the overall best-fitting model at $\logtyr=9.100$ and $Z=0.035$ (marked with a diamond). The second, third and fourth panels show the values of \fbin, $\Delta x$ and $\Delta y$ for the same models, illustrating their small variations around the position of the overall best-fit model, and the almost-null values that $\Delta x$ and $\Delta y$ assume at the best-fit position. The last panel shows the derived fraction of very fast rotators (i.e. those with $\omegai>0.7$). This latter panel evinces not only that the best solution has a large fraction of very fast rotators, but also that solutions with small fractions of fast rotators are just found for age-metallicity values far from the best-fitting one.}
    \label{summary_ngc419}
\end{figure*}

\begin{figure*}
    \centering
    \includegraphics[width=\textwidth]{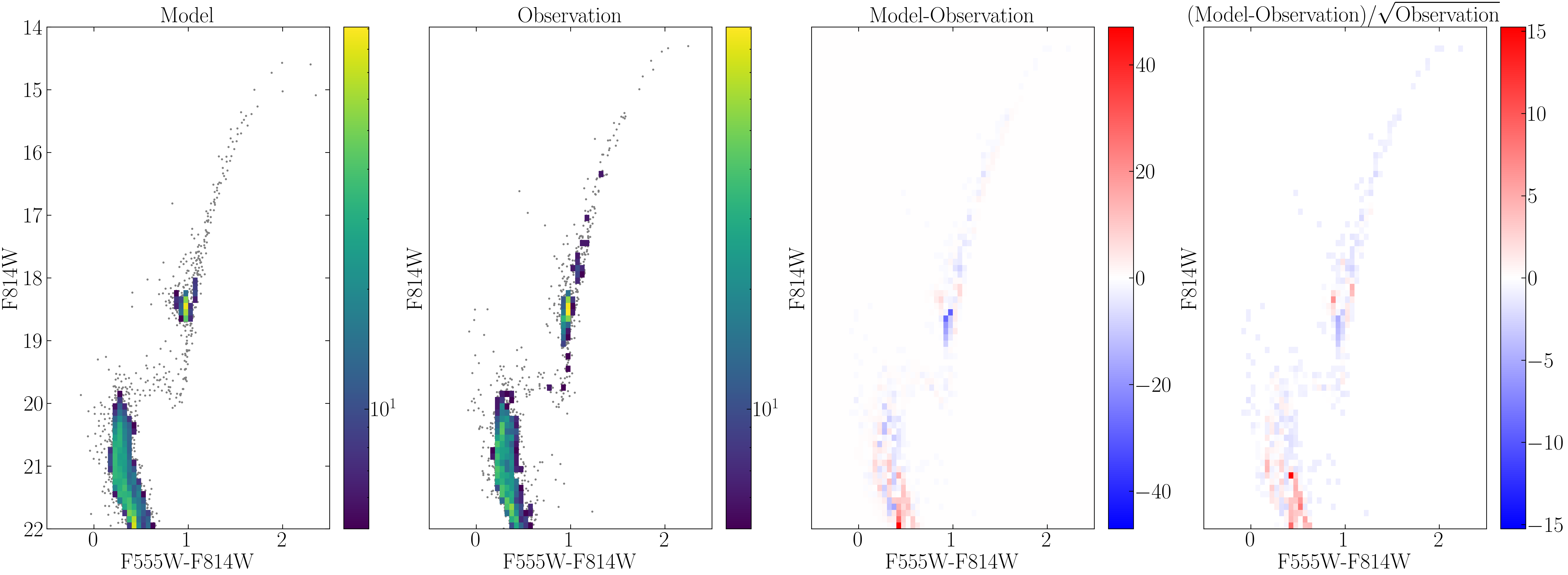}
    \caption{The best-fitting model for NGC 419. From left to right: the Hess diagram of the best-fitting model corresponding to a metallicity of $Z=0.0035$ and age $\logtyr=9.100$; the Hess diagram of the observations; the residuals (model minus observations); and the normalized residuals, i.e. the ratio between the residuals and the Poisson noise of the observations at every CMD bin. The colourbar represents the counts across the Hess diagrams.}
    \label{result419}
\end{figure*}

\paragraph*{NGC 419} 
The results are summarized in Fig.~\ref{summary_ngc419} and Table~\ref{fittingparam}.
The highest likelihood, which corresponds to the lowest $- \ln (\text{PLR})$ value, is highlighted and located at $\logtyr=9.100$ and $Z=0.0035$. The middle-left panel shows the binary fraction determined at each point of the grid. All solutions point to a low binary fraction, with the highest likelihood model having $\fbin=0.140$. The middle-right and bottom-left panels show the CMD shifts, and inform about the changes on the assumed extinction and distance to the cluster. The model chosen as the overall best-fit presents small values for both quantities, leading to very small changes to the initially assumed values. Finally, the bottom-right panel presents the fraction of fast rotators. This fraction is computed dividing the sum of the $a_j$ coefficients for the bins of $\omegai =$ 0.7-0.8, 0.8-0.9, 0.9-0.99 by the total sum of all the 
$a_j$ coefficients. For NGC 419, it appears that the best-fitting model has a very high fraction of fast rotators ($f_{\textrm{highrot}}=0.96$).

The Hess diagram of the best-fitting model is represented in Fig.~\ref{result419}, together with the Hess diagram of the observation, the residuals and the normalized residuals. To avoid areas of the CMD with low completeness levels, we have decided to apply a mask to fit the model to the data only within $14< \rm{F814W} < 22 $ and $-0.5 < \rm{F555W-F814W} < 2.5$. Looking at the first panel, it appears that the analysis provides a very good fit at the level of the MSTO, where the residuals are close to zero. Moreover, the model exhibits a fairly extended RC region, although some negative residuals are present and are clearly visible in the third panel.
This may suggest that either relying solely on stellar rotation is not sufficient to replicate the full extension of the RC in NGC 419, as also found by \cite{Dresbach2023}, or that the adopted models might be underestimating the amount of RC stars. 

\begin{figure*}
    \centering
    \includegraphics[width=\textwidth]{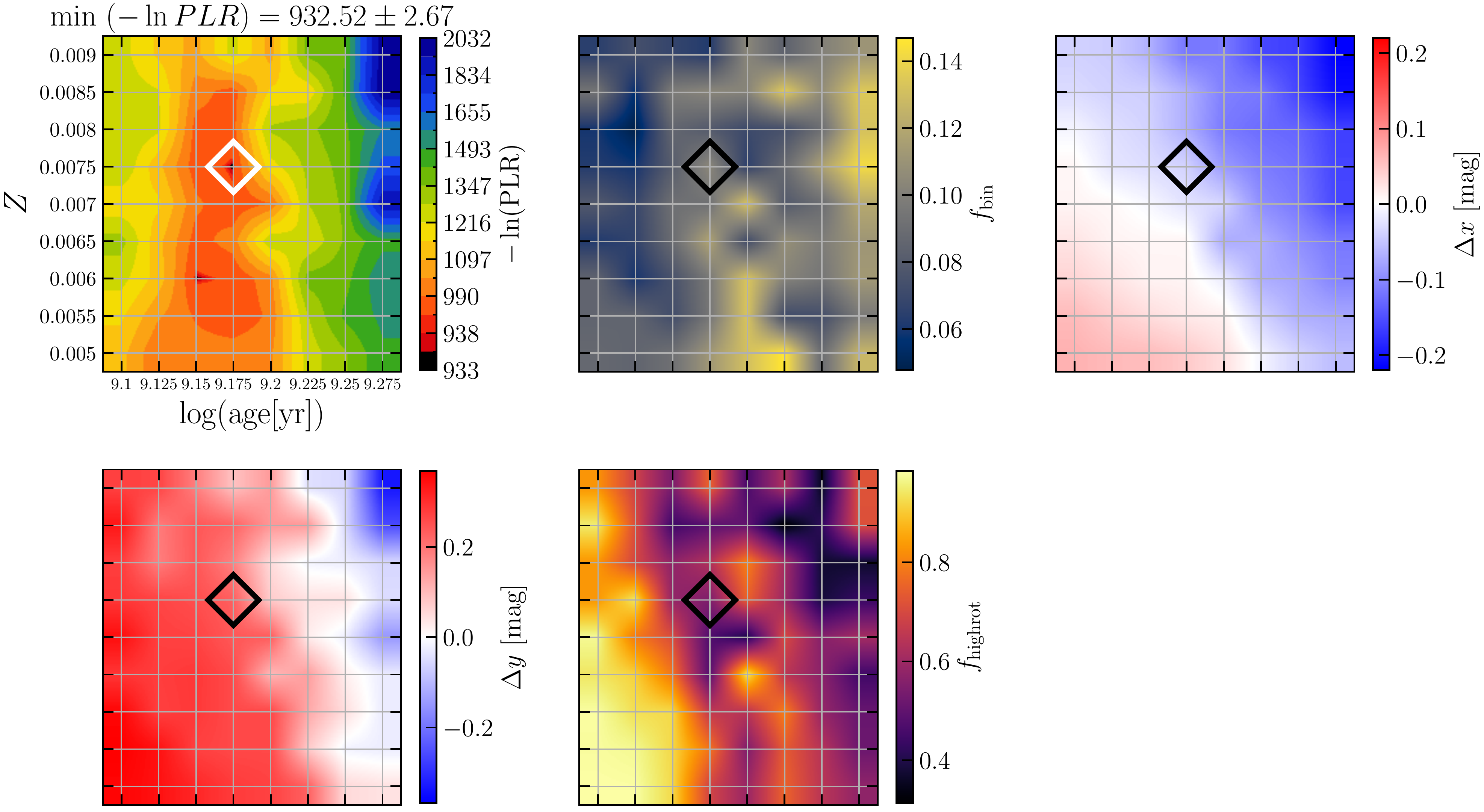}
   \caption{Similar to Fig.~\ref{summary_ngc419} but for NGC 2203.}
    \label{summary_ngc2203}
\end{figure*}

\begin{figure*}
    \centering
    \includegraphics[width=\textwidth]{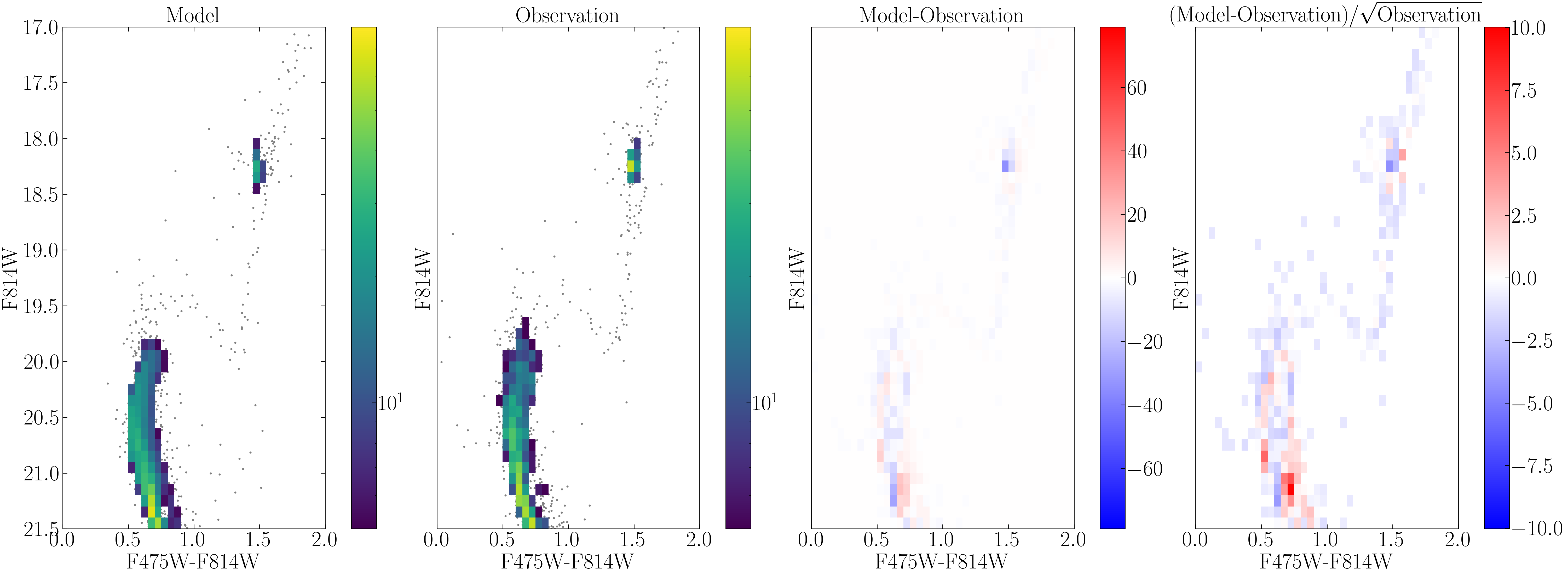}
    \caption{Similar to Fig.~\ref{result419} but for NGC 2203.}
    \label{result2203}
\end{figure*}

\paragraph*{NGC 2203} 
The results are presented in Fig.~\ref{summary_ngc2203}. 
The model with the highest likelihood corresponds to the grid point with $\logtyr=9.175$, metallicity $Z = 0.0075$ and has a relatively small binary fraction, $\fbin=0.104$. The colour shift of the highest-likelihood model is small, while the shift in magnitude is slightly higher than the one for NGC 419. With a fraction of $f_{\textrm{highrot}}=0.55$, NGC 2203 exhibits the lowest frequency of fast rotators among all four clusters.
For the same reasons as NGC 419, a mask was applied, limiting the fitted region to $17< \rm{F814W} < 21.5$ and $0 < \rm{F475W-F814W} < 2$. Fig.~\ref{result2203} displays the Hess diagram of the best-fitting model with $Z=0.0075$ and $\logtyr=9.175$. 
The comparison between the best-fitting model and the observation suggest that the fit is overall very good, and is supported by the relatively low residuals. The MSTO is well reproduced, and similar to NGC 419, there is a negative residual at the location of the RC. However, unlike NGC 419, NGC 2203 does not feature a prominent double RC, which could be explained by its older age. This could further suggest that our models do not fully represent the amount of RC stars. 

\begin{figure*}
    \centering
    \includegraphics[width=\textwidth]{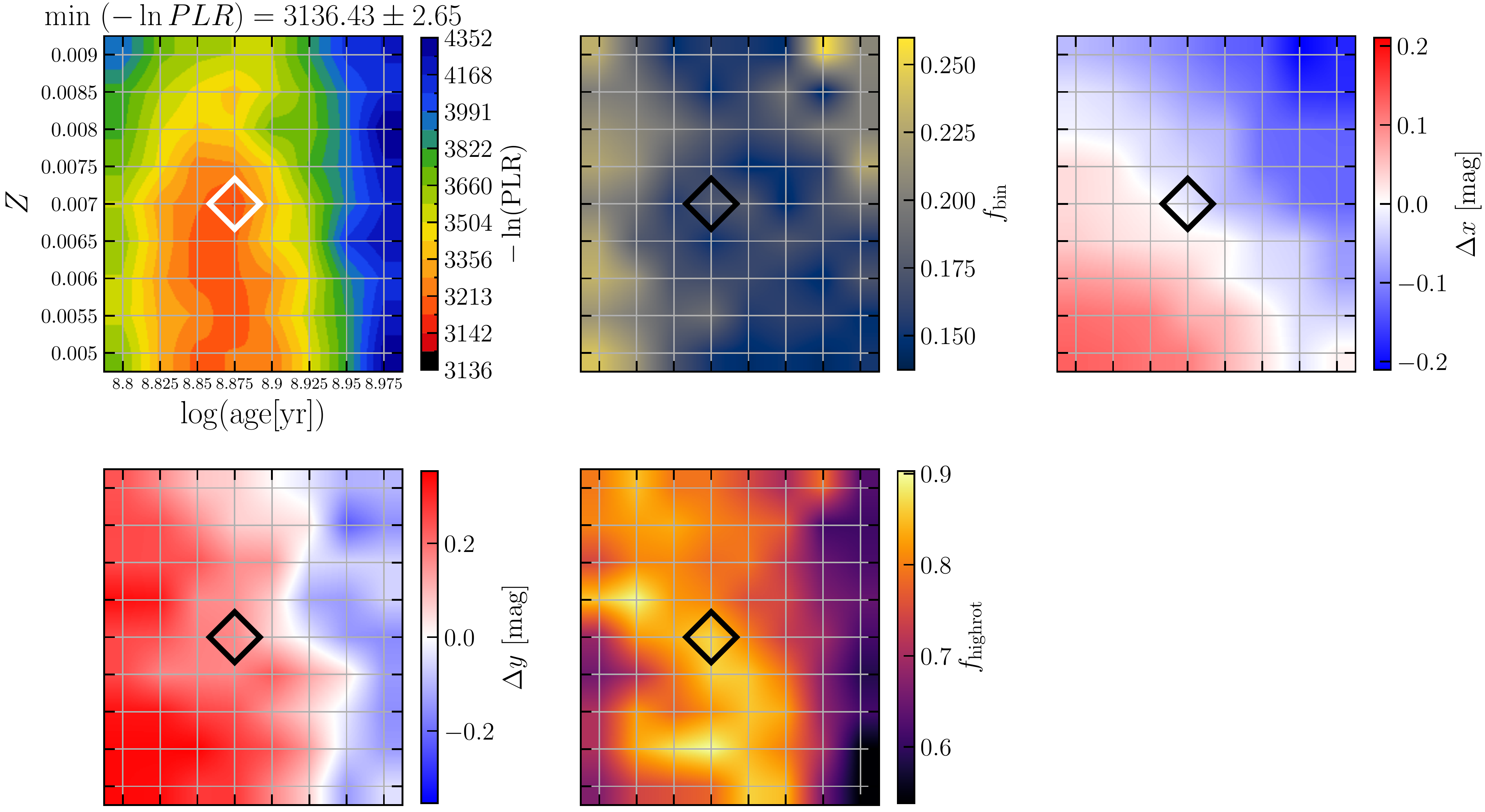}
   \caption{Similar to Fig.~\ref{summary_ngc419} but for NGC 1831.}
    \label{summary_ngc1831}
\end{figure*}
\begin{figure*}
    \centering
    \includegraphics[width=\textwidth]{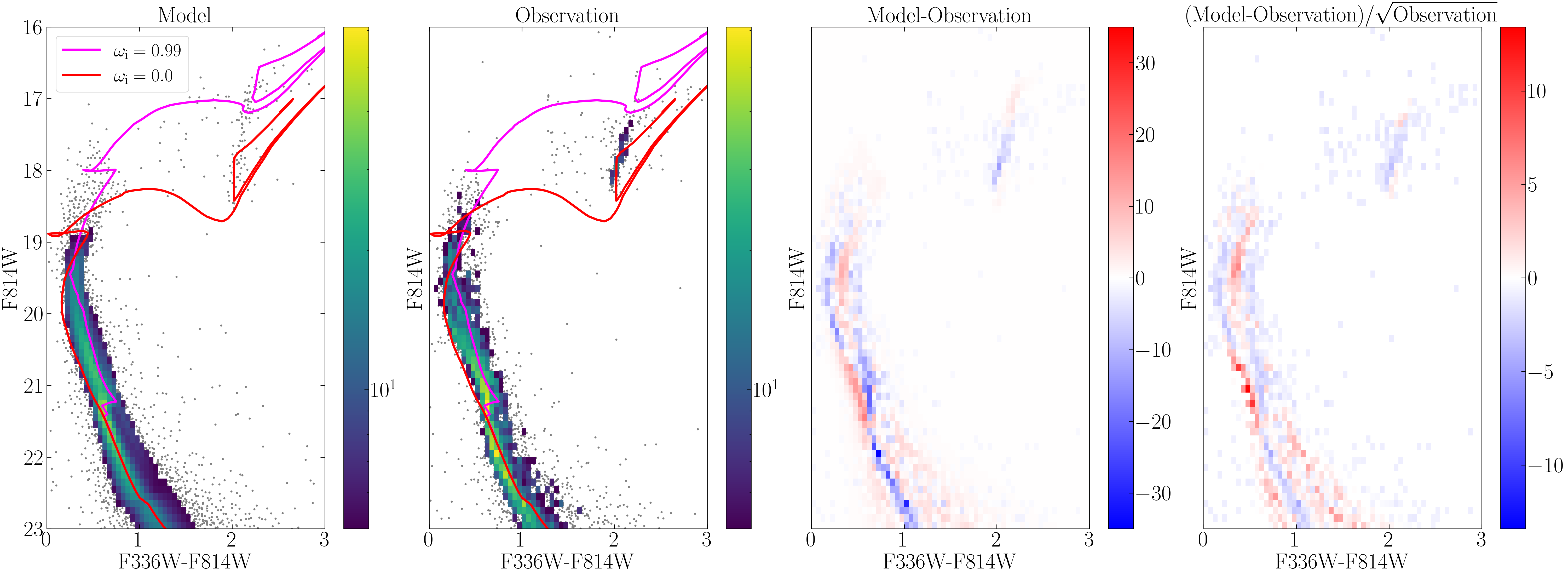}
    \caption{Similar to Fig.~\ref{result419} but for NGC 1831.
    The overplotted \textsc{parsec} isochrones correspond to $\omegai = 0.0$ (in red) and $\omegai = 0.99$ (in magenta), both computed with an inclination $i = 90\degree$.
    }
    \label{1831_iso}
\end{figure*}
\begin{figure*}
    \centering
    \includegraphics[width=\textwidth]{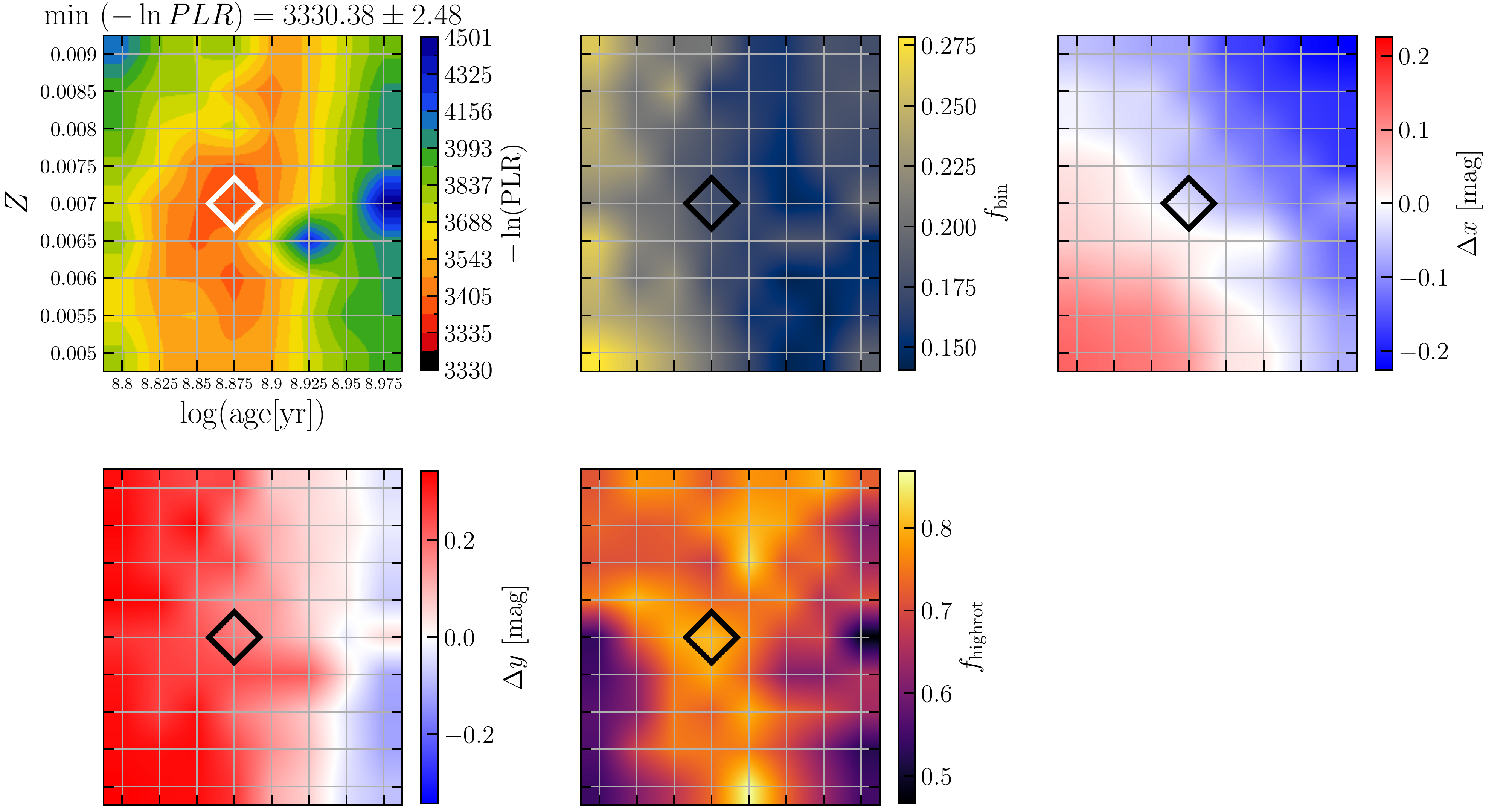}
   \caption{Similar to Fig.~\ref{summary_ngc419} but for NGC 1831 without kink.}
    \label{summary_ngc1831_mask}
\end{figure*}
\begin{figure*}
    \centering
    \includegraphics[width=\textwidth]{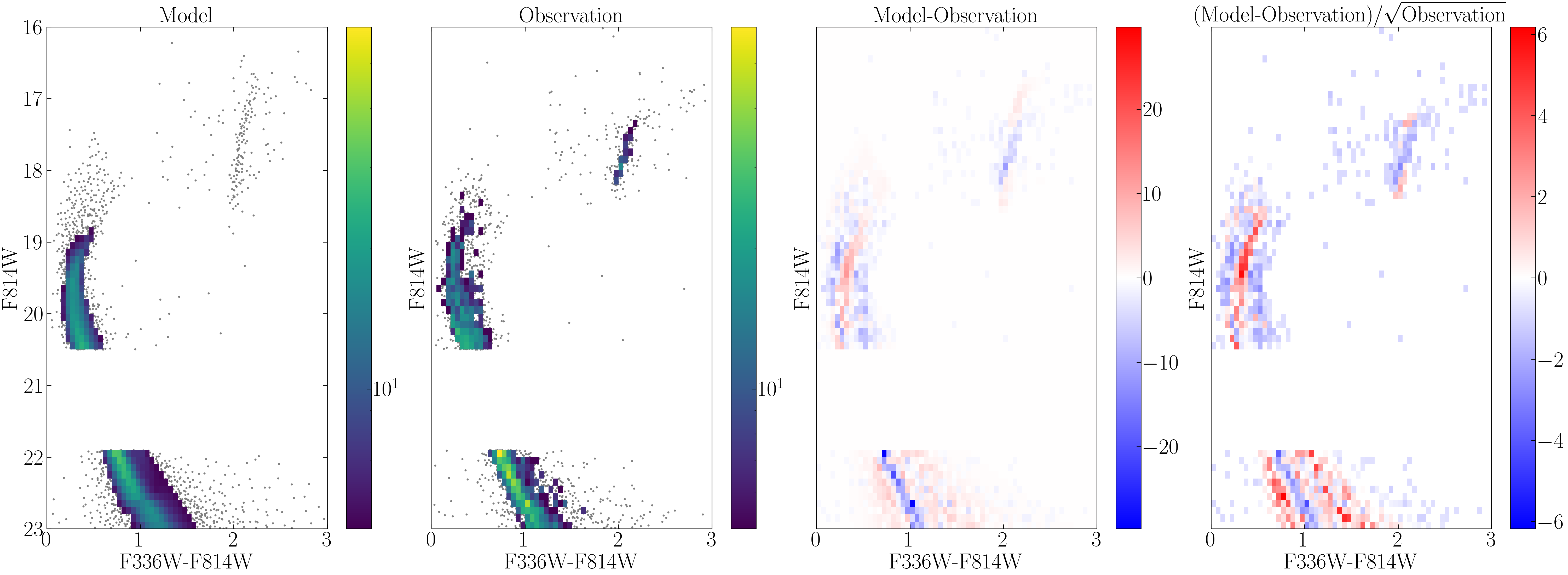}
    \caption{Similar to Fig.~\ref{result419} but for NGC 1831 with a mask applied at the position of the kink. 
    }
    \label{result1831}
\end{figure*}

\paragraph*{NGC 1831}
The analysis was initially analogous to the previous two clusters, with a mask used to select the region of interest and limiting the fit to magnitudes $23< \rm{F814W}< 16$, and colours $0.0 < \rm{F336W-F814W} < 3$. The summary plot obtained from this fit is shown in Fig.~\ref{summary_ngc1831}. 
The best-fitting model has metallicity $Z=0.007$ and age $\logtyr=8.875$.
This is the cluster showing the highest fraction of binaries among all with $\fbin=0.173$. The shifts in colour and magnitude of the best-fitting model are relatively small, and the fraction of fast-rotating stars is very high.
Fig.~\ref{1831_iso} shows the Hess diagram of this model, together with the diagram of the observation, the residuals and the normalized residuals. A feature appears in the diagram of the model between $\rm{F814W} = 21$ and $\rm{F814W} = 21.5$, consisting of an excess of stars bluer than the MS, while such stars are not observed. To infer what kind of stars populate this region, we overplot two \textsc{parsec v2.0} isochrones respectively with $\omegai = 0.0$ (in red) and $\omegai = 0.99$ (in pink), both with inclination $i = 90\degree$. We conclude that the stars populating this region of the model are slowly rotating stars. Moreover, this feature is located close to the level of the MS kink.
In Sect.~\ref{ngc1831}, we explained how the MS kink is artificially introduced when computing \textsc{parsec v2.0} isochrones \citep{PARSEC}, with a gradual growth of rotation along an initial mass interval of $\Delta M_i=0.3\,\Msun$. Hence, the observed overabundance of stars, missing in the observations, might be caused by the transition from non-rotating to rotating stars as implemented in \textsc{parsec v2.0}, and could, therefore, compromise the fit. For this reason, we decided to apply a different mask to keep just the region of the diagram with magnitude ranges $16 < \rm{F814W} < 20.5$ and $22< \rm{F814W} < 24$, and colour range $-0.5 < \rm{F336W-F814W} < 3$, excluding the MS kink.

The likelihood indicates that the model that provides the best-fit of the CMD of NGC 1831 corresponds to $Z=0.007$ and $\logtyr=8.875$, hence the exclusion of the kink has no effect on the values of age and metallicity of the best-fitting model.  The best-fit value of the fraction of binaries is not affected by the change of the mask, while the shift in magnitude and colour is slightly bigger, resulting in a higher value for the distance and a smaller value of the V-band extinction with respect to the fit including the kink.
When comparing the bottom right panel of Fig.~\ref{summary_ngc1831} with Fig.~\ref{summary_ngc1831_mask}, it appears that the fraction of fast rotators is slightly greater in the former, where $f_{\textrm{highrot}}=0.87$. However, even in the latter, this fraction is remarkable ($f_{\textrm{highrot}}=0.80$).
One may observe that the likelihood distribution in Fig.~\ref{summary_ngc1831_mask} exhibits two local maxima of $- \ln (\text{PLR})$ not far from the location of the minimum. Clearly, something caused the solutions at these two points to deviate significantly. While this might seem concerning, excluding these two points results in a well-behaved distribution with only minor noise. Nevertheless, we choose to present the maps as they are since these two deviant points do not change any of our conclusions. Increasing the resolution of the age-metallicity map could be considered, but it would not provide any meaningful new information, as it would only capture interpolations between evolutionary tracks, which are computed with a fixed spacing in both initial mass and metallicity.
The Hess diagram of the new best-fitting model, the Hess diagram of the observation, the residuals and the normalized residuals are plotted in Fig.~\ref{result1831}. Looking at the two models we obtained for the two cases (first panels of Figs.~\ref{1831_iso} and \ref{result1831}), it appears that the quality of the fit is very similar in both cases.
In fact, the MSTO of the model is slightly shifted towards higher luminosity and hotter temperature with respect to the MSTO of the observation, while the stars belonging to the red clump are distributed along a wider range of luminosities than expected. 
One possible reason could reside in the assumption of having stars with an isotropic distribution of inclinations. As we said, if a rotating star is viewed pole-on it will appear brighter and hotter with respect to the same star viewed equator-on. Therefore, the model might be overestimating stars with a pole-on configuration in comparison to the observations, and the cluster's inclination distribution may not be isotropic. In fact, the inclination distribution may even be characterized by a single value of the inclination angle, due to a possible spin alignment of the stars. It could be interesting, in the future, to assume a different distribution for the inclination angle.

\begin{figure*}
    \centering
    \includegraphics[width=\textwidth]{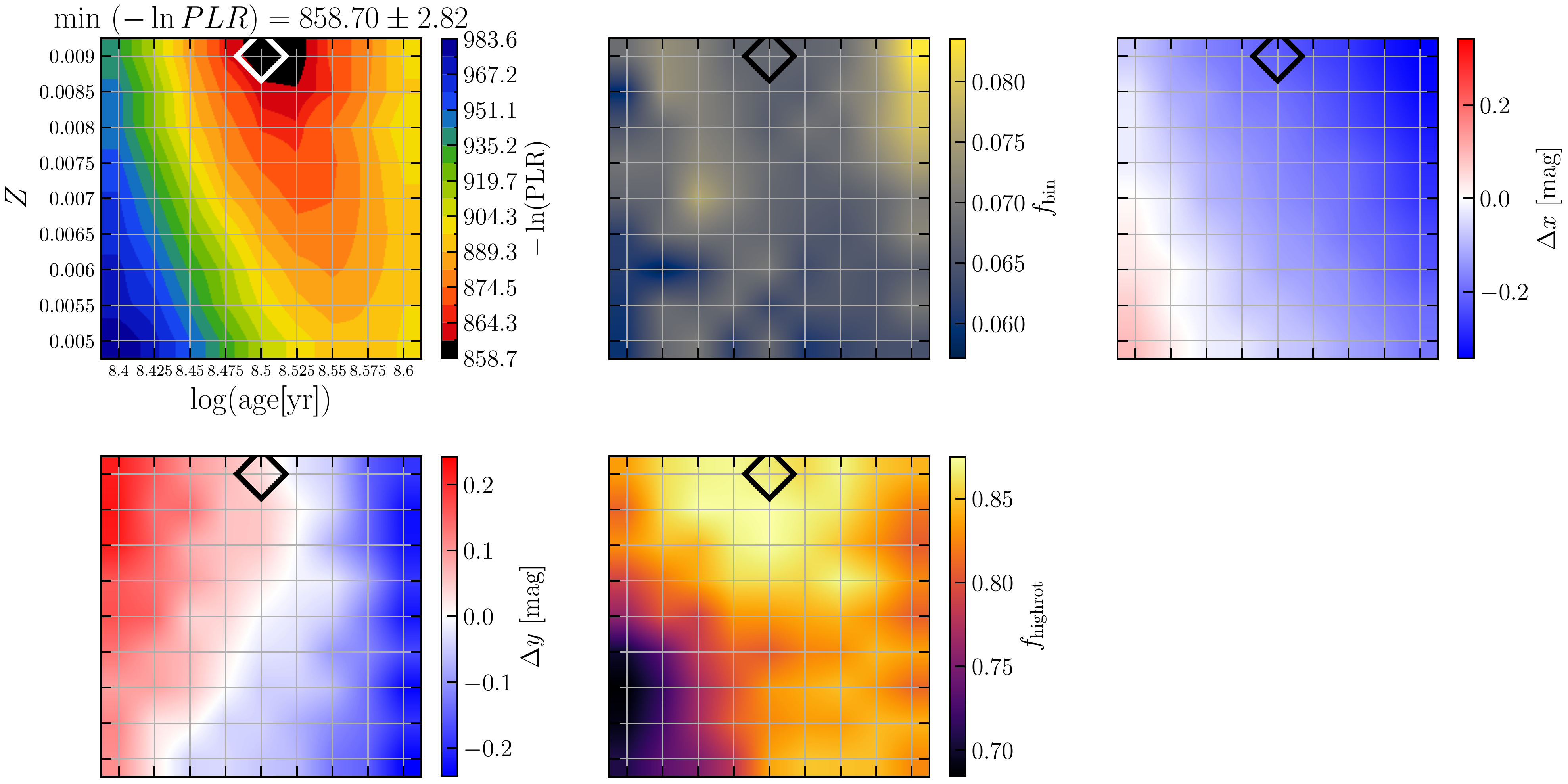}
   \caption{Similar to Fig.~\ref{summary_ngc419} but for NGC 1866.}
    \label{summary_ngc1866}
\end{figure*}
\begin{figure*}
    \centering
    \includegraphics[width=\textwidth]{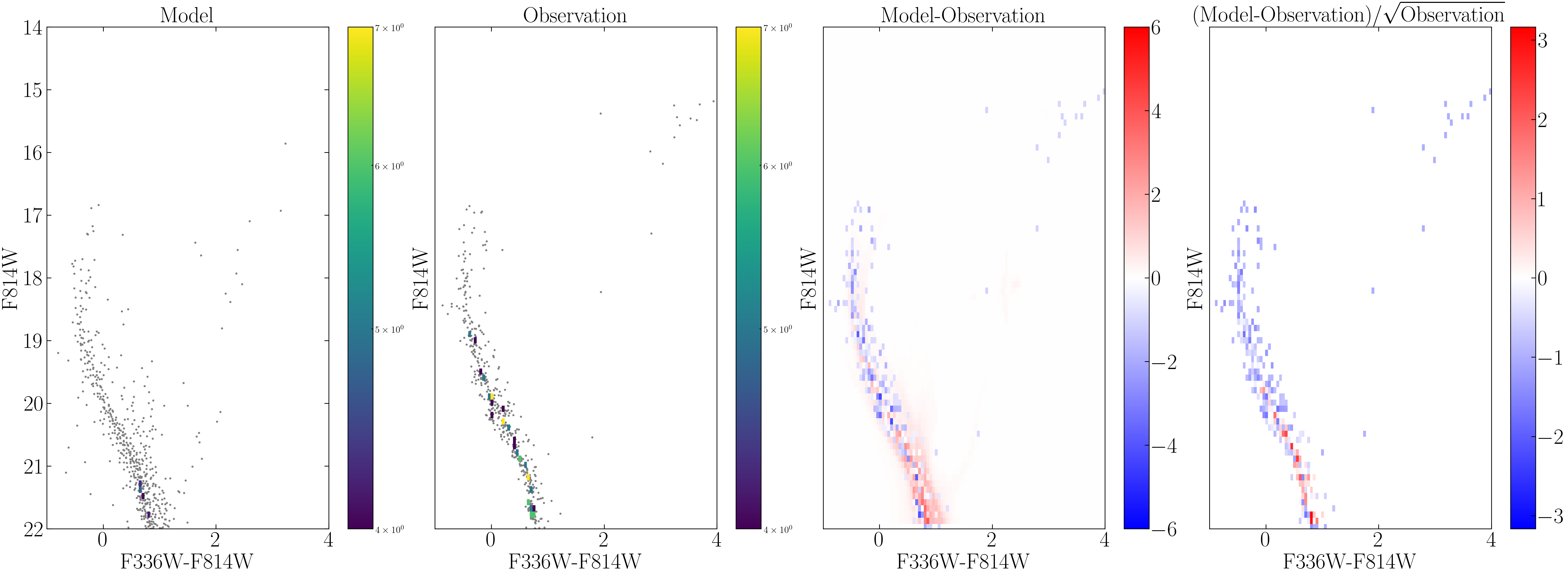}
    \caption{Similar to Fig.~\ref{result419} but for NGC 1866.}
    \label{result1866}
\end{figure*}

\paragraph*{NGC 1866} 
The results are shown in Fig.~\ref{summary_ngc1866}. 
The model with the highest likelihood is the one with $\logtyr=8.5$ and $Z=0.009$ and the binary fraction of this object is $\fbin=0.067$, which is the lowest among all the clusters. Contrary to the other cases, the shift in magnitude is very small, while the colour shift seems to be larger. As for NGC 419 and NGC 1831, the best-fitting model shows a very high value for the fraction of fast-rotating stars, being $f_{\textrm{highrot}}=0.86$.

The Hess diagram of the best-fitting model is represented in Fig.~\ref{result1866}, together with the Hess diagram of the observation, the residuals and the normalized residuals. In this case, a mask was applied to cut for $14< \rm{F814W} < 22 $ and $-1.0 < \rm{F336W-F814W} < 4$. The first thing to notice is that, unlike other clusters, there were very few data available in this case. Specifically, the circular area centered on the cluster with a radius of $29.3$ arcsec contains a total amount of $512$ stars, which is much lower than the other clusters in this study and could potentially affect the statistical analysis.
Unlike the study by \cite{Gossage2019}, where the fit of the eMSTO in NGC 1866 assuming a pure rotation scenario was the least accurate among the three possible scenarios, the best-fitting model in this study provides a quite good reproduction of the eMSTO in NGC 1866.
Moreover, the model shows less stars in the blue part of the MS, especially for $\rm{F814W} \lesssim 21$, hence not fully reproducing the peculiar split MS of NGC 1866. 

The split MS is often linked to the presence of two populations: a slow-rotating group occupying the blue MS, and a portion of fast rotating stars in the red part of the MS. However, looking at the distribution of stellar rotational velocities in NGC 1866, as shown in the top right panel of Fig.~\ref{rotationdistribution}, we clearly see that the presence of stars rotating with $\omegai<0.6$ is negligible, while only fast and extreme rotators seem to populate this cluster, even though an arbitrary fraction of slow rotators is allowed during the fitting.
Either the code encountered challenges in replicating the split MS,  
or the presence of a split MS in this cluster might not be linked to stellar rotation. The Bayesian analysis performed by \cite{Costa2019Cepheids} on NGC 1866 indicates that the blue part of the MS is populated by slowly rotating stars, while fast rotators populate the red MS. 
Given that slow-rotating stars of the blue MS are generally lower in number with respect to the red MS fast-rotating population, and more so in the central regions with respect to the outer ones, it is plausible that our analysis might have exclusively captured the red MS's fast rotators, at a specific age, and that slow rotators of that same age simply do not fit the remaining CMD features. 

We recall that \cite{Costa2019Cepheids} found an age separation of $\sim$112 Myr between the blue MS stars and the stars belonging to the red MS, while in this work, we adopt a pure rotation scenario, assuming that all PMs have the same age. Hence, this assumption could be the reason for the discrepancy between our results and the literature. NGC 1866 may indeed host populations of multiple ages.

\cite{Costa2019Cepheids} also conclude that 4/5 of the Cepheids studied in NGC 1866 evolved from a slowly rotating population, and only one was initially fast rotating. These numbers are in apparent conflict with the results derived in the present paper.

For future works, it could be interesting to perform the same analysis on other clusters that exhibit a split MS and see whether these issues arise in those cases as well.

\subsection{Discussing the uncertainties on age and metallicity}\label{uncertainties}

In order to discuss the uncertainties in the age and metallicity values obtained for these four clusters, we created a $- \ln (\text{PLR})$ color map such that the first three levels represent $1 \times \sigmaPLR$, $2 \times \sigmaPLR$, and $3 \times \sigmaPLR$, as shown in the previous subsection. For clusters NGC 419 and NGC 1831 (both with and without the MS kink mask), we find no other model for which the value of the $- \ln (\text{PLR})$ falls within $1 \times \sigmaPLR$ from the best-fitting model, as shown in Figures~\ref{summary_ngc419}, \ref{summary_ngc1831}, and \ref{summary_ngc1831_mask}. Additionally, in all three cases, the figures indicate that no other model in the grid has a $- \ln (\text{PLR})$ value even within $3 \times \sigmaPLR$ from the best-fitting one. In these instances, we set an upper limit for the uncertainties in age and metallicity, equal to half of the grid step. Thus, for NGC 419 and NGC 1831, the uncertainties are $\pm 0.0125$ for age and $\pm 0.0003$ for metallicity.

In contrast, for NGC 1866, there is another grid point, besides the best-fitting model, where the $- \ln (\text{PLR})$ value lies within $1 \times \sigmaPLR$ of the best-fitting model's value, as shown in Fig. \ref{summary_ngc1866}. This point corresponds to $\logtyr = 8.525$ and $Z=0.009$. Here, the uncertainty for both parameters is defined by the interval covering solutions within $1 \times \sigmaPLR$. Given that the best-fitting model has $\logtyr = 8.500$ and $Z=0.009$, the uncertainties are $\pm 0.0250$ for age and $\pm 0.0003$ for metallicity.

\begin{figure*}
    \centering
    \includegraphics[width=\textwidth]{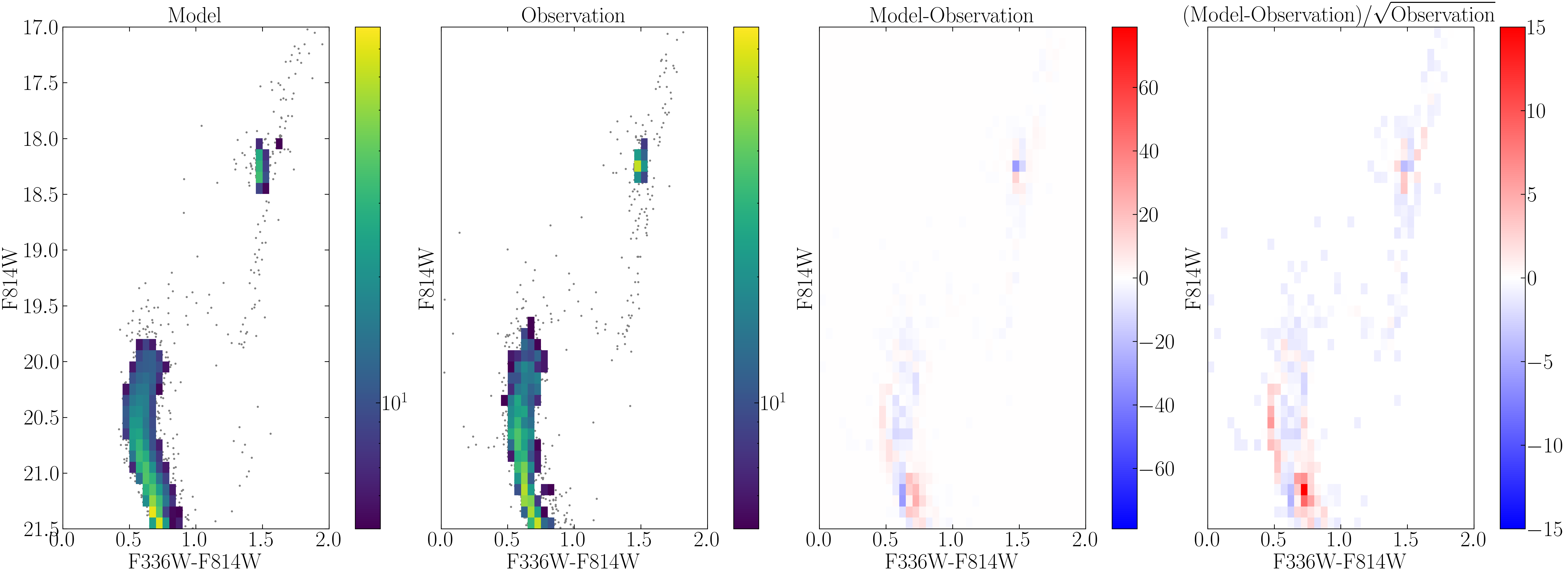}
    \caption{The $\logtyr=9.150$ and $Z=0.006$ alternative model for NGC 2203.}
    \label{result2203_2}
\end{figure*}

Finally, for NGC 2203, another model’s $- \ln (\text{PLR})$ value falls within $1 \times \sigmaPLR$ of the best-fitting model's value. However, this case needs more investigation. In fact, as shown in Fig.\ref{summary_ngc2203} and Table \ref{fittingparam}, this model is located in a different region of the grid, at $\logtyr = 9.150$ and $Z=0.006$. For this reason, we decided to show this alternative model as well in this paper. 
We further investigated this alternative solution and, in Fig.~\ref{result2203_2}, we present the Hess diagram of this model along with the observations, residuals, and normalized residuals. Comparison shows that this model provides, apparently, a poorer fit than the best-fitting model in Fig.~\ref{result2203}, with a fainter TO point with respect to the observation and higher residuals along the MS compared to our best-fitting case; nonetheless, the likelihood values are similar. The distribution of stellar rotational velocity reveals a substantial fraction of fast rotators (\( f_\mathrm{highrot} = 0.91 \)), which is higher than in the first solution, as indicated by the colormap in the final panel of Fig~\ref{summary_ngc2203} and the distribution shown in the bottom-right panel of Fig~\ref{rotationdistribution}. In contrast to NGC~1866, in this case we consider these models as two separate solutions, each one with uncertainties on age and metallicity given by half of the grid step.
\subsection{Distribution of rotational velocities}\label{rotatiodistr}
\begin{figure*}
    \centering
    \includegraphics[width=\textwidth]{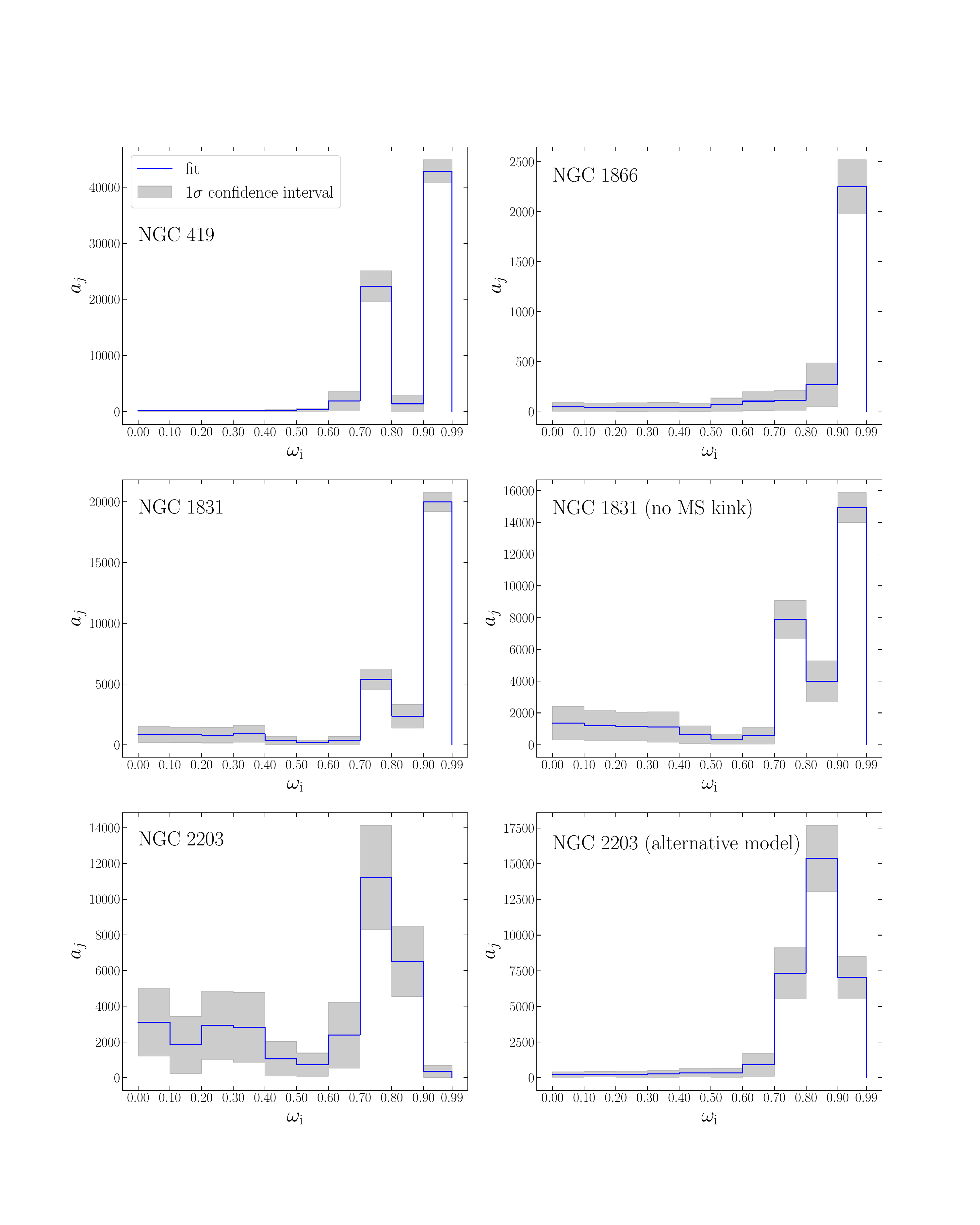}
    \caption{Distribution of stellar rotational velocities in NGC 419 (top left panel), NGC 1866 (top right panel), NGC 1831 with kink (middle left panel), NGC 1831 without kink (middle right panel) and NGC 2203 (bottom left panel). The bottom right panel represents the rotational velocity distribution of the alternative model for NGC 2203. Grey boxes represent the $1\sigma$ confidence interval.}
    \label{rotationdistribution}
\end{figure*}
The best-fitting values of the $a_j$ coefficients of Eq.~\ref{model}, along with the $1\sigma$ confidence interval, are plotted in Fig.~\ref{rotationdistribution} for every cluster of this study. These plots represent their distribution of stellar rotational velocities. The top left panel represents the outcome of the analysis for NGC 419 and shows that this cluster harbors stars with $\omegai\geq0.7$, with a tiny fraction of stars with $0.8\leq\omegai\leq0.9$ and a negligible amount of stars with $\omegai<0.7$. The majority of the cluster's stars has $0.9\leq\omegai\leq0.99$.
This suggests that the morphology observed in its CMD, with its distinct features, can be explained by the presence of extreme rotators among its stellar population.

Contrary to NGC 419, the bottom left panel shows that NGC 2203 does not include stars with break-up velocity, on the other side it seems to be mainly populated by stars with $0.7\leq\omegai\leq0.9$. This cluster stands out as the only cluster where a discernible group of stars exhibits $\omegai \lesssim 0.5$, although the $1\sigma$ confidence intervals are larger than the other cases. 

In the middle left panel, we plot the rotation distribution of NGC 1831 obtained from the fit including the MS kink, while the middle right panel shows the distribution following the fit with the updated mask excluding the kink. In both cases, the majority of the stars are very close to the break-up velocity, having $0.9\leq\omegai\leq0.99$. Moreover, this cluster is also populated by moderately fast rotators with $0.7\leq\omegai\leq0.8$, similar to the case observed in NGC 419. The fit excluding the MS kink suggests that stars are more evenly spread across the bins where omega $\omegai \geq 0.7$, compared to the results obtained from the fit that includes the MS kink.
\cite{Correnti2021} studied this cluster comparing its CMD morphology with Monte Carlo simulations involving synthetic star clusters consisting of multiple populations with different ages or a single population with a range of rotational velocities, using the Geneva SYCLIST isochrones. They inferred that the morphology of NGC 1831 could be fully explained, within the context of a pure rotation scenario, assuming a bimodal distribution for the rotating stars, with $\sim 40$ per cent of slow rotators and the remaining $\sim 60$ per cent being fast rotators. On the other side, both our fits for NGC 1831 suggest a higher fraction of fast and extreme rotators with respect to \cite{Correnti2021}.

The distribution of rotational velocities in NGC 1866 is represented in the upper right panel. As already anticipated in Sect.~\ref{bestfitmodels}, we clearly see that the presence of stars rotating with $\omegai<0.9$ is negligible, while only extreme rotators seem to populate this cluster.

Finally, the bottom right panel shows the rotational velocity distribution of the alternative model proposed for NGC 2203. As anticipated in Sect~\ref{uncertainties}, the alternative model provides a distribution peaked at significantly higher velocities compared to the first model, with $\omegai \geq 0.7$.

\section{Conclusions}\label{conclusion}
In this paper, we delve into a comprehensive study of the stellar rotation distribution of four target clusters of the MCs through the application of models that are complete in all evolutionary phases and include rotation.
The analysis concerns four clusters: NGC 419 of the SMC and NGC 2203, NGC 1831 and NGC 1866 of the LMC. All these objects present all the interesting features associated with rotation in their CMDs. Using isochrones derived
from \textsc{parsec v2.0} stellar tracks \citep{PARSEC}, we generate distinct stellar populations, each covering a different range of initial rotation rate \omegai. These stellar populations are referred to as `Partial Models' (PMs), and they are generated assuming an isotropic distribution of the inclination angle. The final best-fitting model is obtained applying a method that can be described as "CMD fitting", in which the observed CMD is compared to a model CMD made by the linear combination of the PMs. This is done with \texttt{sfhfinder} \citep{Mazzi}, which allows us to derive both the best-fitting solution and the confidence intervals of the fitted parameters. 
These models are generated over a grid of age and metallicity values, resulting in the corresponding amount of best-fitting models M given by Eq.~\ref{model}. 
Given the Poissonian nature of the distribution of our data, we chose to compute and minimize Eq.~\ref{PLR} as the most suitable approach to identify the best model among those several best-fitting models. 
The age and metallicity values are determined from the grid point with the maximum likelihood as the result of a model selection process, as described in Sect.~\ref{simulations}. For each studied cluster, comparing the likelihoods in all simulated models shows that, in most cases, no other models -- or at most one in the case of NGC 1866 and NGC 2203 -- approach the likelihood of the best-fitting model within $1 \times \sigmaPLR$. This indicates that the selected model offers the closest match to the observed data and provides the best estimates for the cluster’s age and metallicity. Moving forward, future analyses might refine these results further by incorporating simultaneous fitting of all key parameters, including age and metallicity.

The age of NGC 419 is estimated to be $\sim 1.3$ Gyr, which is younger with respect to the assumed value of 1.45 Gyr inferred by \cite{Goudfrooij2014} who used the non-rotating isochrones of \citet{Marigo08}. Similarly, NGC 1831 appears to be $\sim 750$ Myr old, which is 50 Myr younger than the $\sim 800$ Myr age estimated by \cite{Correnti2021} using the Geneva SYCLIST isochrones. Conversely, the best-fitting model for NGC 2203 perfectly matches the value of 1.5 Gyr reported by \cite{Goudfrooij2014}. Finally, NGC 1866 is the only cluster of the analysis that is older with respect to the literature \citep[$\sim 250$ Myr;][]{Goudfrooij2018}, with a best-fit age of $\sim316$ Myr.
In general we obtain ages close to the ones from the literature, and importantly in the case of NGC 2203, which has high-quality photometry and a faint completeness limit, the age we find with the best-fitting model is almost exactly equal to previous estimates, while the alternative model suggests a slightly younger age (1.41 Gyr).

In addition to obtaining the distribution of rotational velocities, we chose to employ \texttt{sfhfinder}  \citep{Mazzi} to fit the binary fraction and the shifts in magnitude and colour as well, for each cluster.
The first quantity provides insight into the amount of binary systems with respect to the total number of stars in the cluster. Notably, NGC 1831 exhibits the highest count of binary systems ($f_{\textrm{bin}} = 0.173$) among the objects in this study, while NGC 1866 displays the lowest count ($f_{\textrm{bin}} = 0.067$). The shifts in magnitude and colour, on the other hand, can be used to calculate the changes from the reference values of distance and extinction after the fitting procedure. Indeed, when isochrones are processed before generating stellar populations, we assumed for each cluster a given value for the distance modulus $(m-M)_0$ and V-band extinction $A_{\mathrm{v}}$, that are listed in Table~\ref{tab:paramsMC}. 
The distance and V-band extinction calculated after the fit are presented in Table~\ref{fittingparam}. In NGC 419, the derived distance after the fit is 0.34 kpc closer than the assumed value. For the remaining clusters, the distances are all slightly larger, ranging from an increase of 3.5 kpc for NGC 1866 to a maximum difference of 6.2 kpc obtained for NGC 2203. The V-band extinction values obtained from the fit are generally consistent with the initial assumptions for NGC 419, NGC 2203 and NGC 1831, with a maximum difference of 0.06 for the latter. On the other side, NGC 1866 exhibits a significant difference in V-band extinction. The fit suggests a value of $0.08$\,mag, which is considerably lower than the assumed value of $0.28$\,mag.

The most striking conclusion reached through this process, however, is the very high fraction of fast rotators that we find in all four clusters:
NGC 419 is predominantly populated by stars with $\omegai\geq0.70$, including a very high proportion of stars with $0.90\leq\omegai\leq0.99$, and a negligible number of stars with $\omegai<0.70$. 

Meanwhile, NGC 2203 does not include stars with break-up velocity, but it seems to be mainly populated by stars with $0.7\leq\omegai\leq0.9$. This cluster stands out as the only cluster where a discernible group of stars exhibits $\omegai \lesssim 0.5$, although the $1\sigma$ confidence intervals are bigger than the other cases. On the other side, the alternative model suggests a population of very fast rotating stars, with $\omegai\geq0.70$, including stars with break-up velocity.

The distribution of NGC 1831 shows that, whether we include the MS kink in the fit or not, the majority of its stars are very close to the break-up velocity, having $0.9\leq\omegai\leq0.99$. Moreover, this cluster is also harbors a stellar population of moderately fast rotators with $0.7\leq\omegai\leq 0.8$.
Finally, NGC 1866 only harbors stars with $\omegai\gtrsim 0.90$.

How do these results compare with actual measurements of rotational velocities in star clusters? The few spectroscopic studies to date always include a significant fraction of slowly rotating stars, that seems at odds with our findings. However, the differences are not necessarily significant, since 
spectroscopic measurements always produce the present projected rotational velocity, $v\sin i$, star by star, while our method aims to derive the initial distribution of $\omegai$ by mass fraction of the parent population.
Moreover, spectroscopic measurements tend to avoid the most crowded  central regions of the clusters, especially in the older and more populated ones; those regions are always well sampled in our selected \textit{HST} photometry.
Despite these differences in methods and samples, what is very clear is that spectroscopic studies tend to derive conspicuous fractions of low-$v\sin i$ stars, which apparently are not present among our results. For instance, \citet{Marino2018a} and \citet{bodensteiner23} find populations of slowly rotating stars among the bright main sequence stars (F814W $\lesssim 19$\,mag) of the very young clusters NGC 1818 and NGC 330, respectively, both with ages close to 40 Myr. Such stars have some correspondence with the brightest stars we have in NGC 1866 -- where we instead find fast rotating stars only. 
Indeed, looking at the most recent literature regarding NGC 1866 \citep{Goudfrooij2018,Costa2019Cepheids,Gossage2019}, the distribution of rotational velocities was expected to have a bimodal trend, accounting for the presence of slow-rotating blue MS stars and fast rotators belonging to the red MS. This was also confirmed by the spectroscopic study performed by \citet{Dupree2017} on 29 eMSTO stars of this cluster, whose spectra suggest the presence of two distinct populations of stars: one rapidly rotating, while the other consists of younger, slowly rotating stars.  It is however to be noticed that the observations by \citet{Dupree2017} refer to likely member stars within a 3 arcmin wide field and separated at least 2.5 arcsec from neighbouring bright stars; these criteria result in a spectroscopic sample distributed in the cluster outskirts; indeed they have just one star in common with our \textit{HST} sample (which is limited to a 29.3 arcsec radius from the cluster centre)\footnote{Similarly, the five Cepheids that  suggest the prevalence of slowly-rotating stars in NGC~1866 \citep{Costa2019Cepheids}, are located, on the mean, even farther from the cluster centre than the \citet{Dupree2017} spectroscopic sample.}. And, following the indications by \citet{milone17}, one expects a significant reduction in the fraction of fast rotators as we move to the cluster oustkirts; more specifically, they conclude that the fraction of blue MS stars (interpreted as slow rotators) increases from the central $\sim30$ per cent value to $\sim45$ per cent at radii of 3 arcmin. 

As anticipated in Sect.~\ref{bestfitmodels}, the absence of slow rotators in NGC~1866 resulting from this work may also be due to our choice to assume a unique value of the age for the PMs, while for instance \cite{Costa2019Cepheids} supports the presence of two distinct populations in NGC 1866, with ages of $\sim 176$ Myr and $\sim288$ Myr, respectively. On the other side, more recent results on other young MC clusters \citep[e.g.][]{Cordoni2022} find evidence for a fast star formation. Moreover, it is possible that 
one of our fundamental assumptions -- such as the isotropic distribution of spin axes or the assumption of a single age for both fast and slow rotators -- is not appropriate for this cluster.

Regarding older clusters, a spectroscopic study of NGC 419 was performed by \cite{Kamann2018}. Here they inferred that the eMSTO shows significant rotational broadening, with red MSTO stars rotating faster than bluer stars. Stacking the spectra obtained with the MUSE AOF/GALACSI, they found average $v\sin i$ values of $87 \pm 16$ and $130 \pm 22$\,km s$^{-1}$ for blue and red MSTO stars, respectively.
Additionally, the spectroscopic observations of the 1.5\,Gyr old NGC 1846, do also reveal a conspicuous population of low-$v\sin i$ stars \citep{kamann20}. More specifically, they find a bimodal distribution of $v\sin i$, which is clearly absent from our \omegai\ distributions for clusters of similar age. 
It remains to be determined how serious is this apparent discrepancy -- by means, for instance, of the CMD fitting of these few clusters which have actual $v\sin i$ measurements.

Our findings serve as a cautionary signal, highlighting the need for significant advancements in the development and calibration of rotating stellar models.
To further explore the full range of possibilities, future studies could incorporate alternative stellar evolutionary models with rotation and explore different assumptions in the CMD fitting, such as allowing for an age spread or a preferential orientation of fast rotators. 
Moreover, whether our result is general remains to be determined. Our analysis focused on clusters with the most obvious signatures of fast rotation -- hence finding that the fast rotators are indeed there. Future deep spectroscopic observations of our four clusters, along with those already performed as \citet{Dupree2017} for NGC 1866 and \citet{Kamann2018} for NGC 419, could help clarify the results presented in this work.
Nonetheless, it is clear that such a result should motivate a thorough revision of the stellar evolution models that have been used for decades in the study of MC clusters. The most massive star clusters in the MCs have for long been considered as a crucial calibrator of stellar evolutionary models, but they have, apparently, calibrated the wrong type of models: the non-rotating ones.

In the future, this analysis could be extended to a larger number of stellar clusters of the MCs with different age and metallicity in order to see whether the outcomes observed in this study hold as a general pattern within these two nearby irregular dwarf galaxies. In this perspective, it may be interesting to study more clusters with split MS and check if the analysis faces similar issues like those encountered in fitting NGC 1866. 
Moreover, this work could be expanded to young and intermediate-age clusters within the MW, providing a more comprehensive understanding of the underlying mechanisms shaping the CMDs in a different galactic environment.
 
\section*{Acknowledgements}
The authors wish to honour the memory of Professor Paola Marigo, who passed away on October 20, 2024. 
Her memory will be forever cherished, and her invaluable teachings, passion, and humanity will continue to inspire all of those who had the privilege to meet her.

Support for this project was provided by NASA through grants HST-GO-14688, HST-AR-15023 and HST-AR-13901 from the Space Telescope Science Institute, which is operated by the Association of Universities for Research in Astronomy, Inc., under NASA contract NAS5-26555.
A.M. acknowledges financial support from Bologna University, ``MUR FARE Grant Duets CUP J33C21000410001''.
LG acknowledges an INAF Theory grant.
YC acknowledges support from the National Natural Science Foundation of China (NSFC) No.~12003001.
GC acknowledges partial financial support from European Union—Next Generation EU, Mission 4, Component 1, CUP: C93C24004920006, project ‘FIRES - FIrst stars makEr Simulations'.
GC acknowledges partial financial support from the Agence Nationale
de la Recherche grant POPSYCLE number ANR-19-CE31-0022.
We also acknowledge the Italian Ministerial grant PRIN2022, ``Radiative opacities for astrophysical applications'', no. 2022NEXMP8.

Part of the research activities described in this paper were carried out with contribution of the Next Generation EU funds within the National Recovery and Resilience Plan (PNRR), Mission 4 - Education and Research, Component 2 - From Research to Business (M4C2), Investment Line 3.1 - Strengthening and creation of Research Infrastructures, Project IR0000034 – ``STILES - Strengthening the Italian Leadership in ELT and SKA''.

\section*{Data Availability}
All data used in this paper are available in the \textit{HST} archive. Reduced and processed data might be provided by request to the corresponding author.
\bibliographystyle{mnras}
\bibliography{bibliography}

\appendix
\section{Mock fits}
\label{app:mocks}

    In this section we present examples of fits performed with \texttt{sfhfinder} of mock observations constructed from the partial models of NGC 419 and NGC 1866 (left and right columns, respectively, in Figures~\ref{fig:mocks_1}, \ref{fig:mocks_2} and \ref{fig:mocks_3}). More precisely, we used the partial models with age $\log(t/\text{yr})=9.0$, metallicity $Z=0.004$ and true distance modulus 19.1 for NGC 419. For NGC 1866 we used the model with age $\log(t/\text{yr})=8.525$, $Z=0.0075$, and true distance modulus 18.43. To build the mock observations we have used the same methods described in Section~\ref{methods}, with the same procedures for applying the photometric errors and the incompleteness.

    In each of the figures presented below, the black line (mock) shows the input velocity distribution and the blue line shows the result fo the fit (result). The gray shaded region represents the $1\sigma$ region, while the light gray-blue region approximates the $2\sigma$ confidence interval.

    We first evaluate the performance of \texttt{sfhfinder} on mock observations constructed from a small subset of partial models using the same binary fraction (30\%), color shift ($-0.25$ mag) and magnitude shift ($+0.3$ mag).
    The code can reliably recover the presence of fast rotators (Figures~\ref{fig:mock_419_fast} and \ref{fig:mock_1866_fast}), generally within $2\sigma$ of the input value.
    In the case of slow rotators (Figures~\ref{fig:mock_419_slow} and \ref{fig:mock_1866_slow}) the code distributes some of the weight in the neighboring bins of rotational velocity.
    This is expected, as the models with slow rotation do not differ significantly from each other.
    A potential solution for this aspect would be to increase the resolution of the Hess diagrams, in exchange for fewer stars in each 2D bin of the diagrams.
    If both very slow and very fast rotating models are used to construct the observations (Figures~\ref{fig:mock_419_two} and \ref{fig:mock_1866_two}), the weight of the fast rotating partial model approximates the input value, while the weight of the slow rotating one is spread in the neighboring bins; nonetheless, the total amount of slow rotators is recovered.

    We repeated the previous test with ten times more stars (Figures~\ref{fig:mock_419_two_morestars} and \ref{fig:mock_1866_two_morestars}), which resulted in the weights for slow and fast rotators being closer to their input values.
    With ten times fewer stars instead (Figures~\ref{fig:mock_419_two_lessstars} and \ref{fig:mock_1866_two_lessstars}) the result degrades: while the presence of the fast rotating model is detected, although at the limit of $2\sigma$, the slowly rotating model is confused with the other slowly rotating ones.
    In the latter case, increasing the number of steps does not improve the results (Figures~\ref{fig:mock_419_two_lessstars_moresteps} and \ref{fig:mock_1866_two_lessstars_moresteps}).

    We also tested the effect of different binary fractions.
    The code has some trouble fitting mock observations with a very high binary fraction (90\%, Figures~\ref{fig:mock_419_fbin90} and \ref{fig:mock_1866_fbin90}).
    This is expected because the presence of a very large fraction of binaries mimics the presence of very fast rotators, at least along the main sequence.
    In the case of a moderately high binary fraction (60\%, Figures~\ref{fig:mock_419_fbin60} and \ref{fig:mock_1866_fbin60}), the code is able to recover the input velocity rotation distribution much better.
    It should be noted that the clusters analyzed in this study have binary fractions that are substantially lower compared to the two examples described above.

    Finally, Figures~\ref{fig:mock_419_random} and \ref{fig:mock_1866_random} show the result of the fits performed with the code on mock observations computed with completely randomized parameters.

    Our main conclusion from the tests presented above is that the code can reliably identify the populations of high-speed rotators in clusters with a low to moderate binary fraction.
    Populations with slow rotational speeds can be masked by being spread by the code into neighboring bins of rotational velocity.
    NGC 1831 and NGC 2203 can indeed have a population of slowly rotating stars that the code was unable to identify.
    However, this population should represent a small fraction of the stars in each cluster.

\begin{figure*}
    \centering
    \begin{subfigure}[t]{0.45\linewidth}
        \centering
        \includegraphics[width=\linewidth]{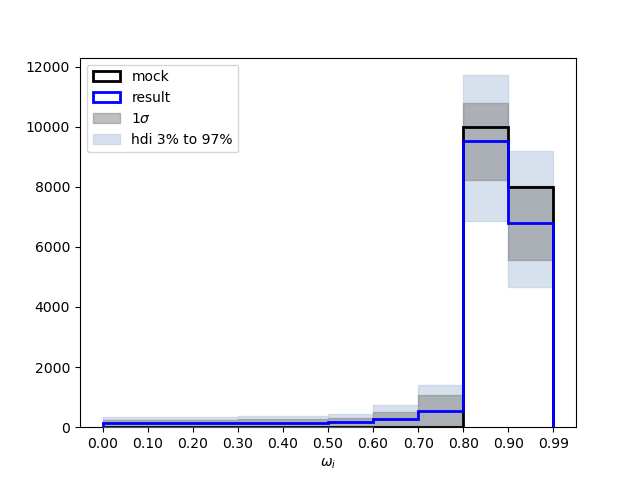}
        \caption{Mock observations made with only the two partial models of NGC 419 with the fastest rotation rate.}
        \label{fig:mock_419_fast}
    \end{subfigure}\qquad
    \begin{subfigure}[t]{0.45\linewidth}
        \centering
        \includegraphics[width=\linewidth]{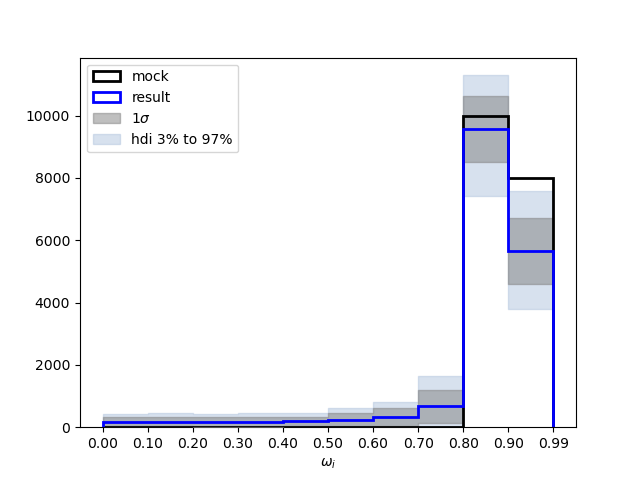}
        \caption{Mock observations made with only the two partial models of NGC 1866 with the fastest rotation rate.}
        \label{fig:mock_1866_fast}
    \end{subfigure}
    \begin{subfigure}[t]{0.45\linewidth}
        \centering
        \includegraphics[width=\linewidth]{Figures/mocks/ngc419/figA1_a.png}
        \caption{Mock observations generated using only the two partial models of NGC 419 with the slowest rotation rate.}
        \label{fig:mock_419_slow}
    \end{subfigure}\qquad
    \begin{subfigure}[t]{0.45\linewidth}
        \centering
        \includegraphics[width=\linewidth]{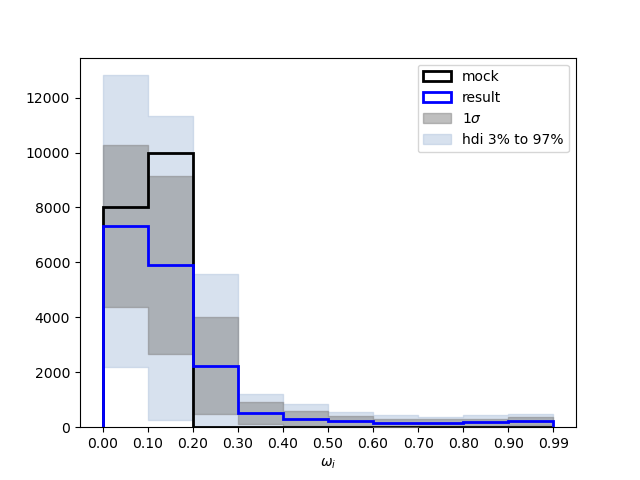}
        \caption{Mock observations generated using only the two partial models of NGC 1866 with the slowest rotation rate.}
        \label{fig:mock_1866_slow}
    \end{subfigure}
    \begin{subfigure}[t]{0.45\linewidth}
        \centering
        \includegraphics[width=\linewidth]{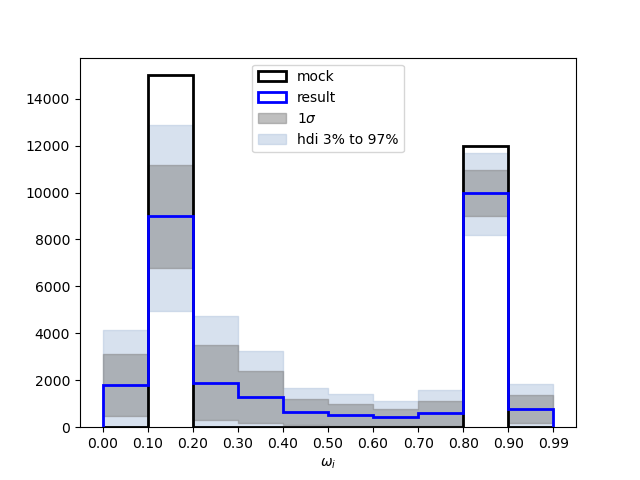}
        \caption{Mock observations made with only one slowly and one fast rotating partial models of NGC 419.}
        \label{fig:mock_419_two}
    \end{subfigure}\qquad
    \begin{subfigure}[t]{0.45\linewidth}
        \centering
        \includegraphics[width=\linewidth]{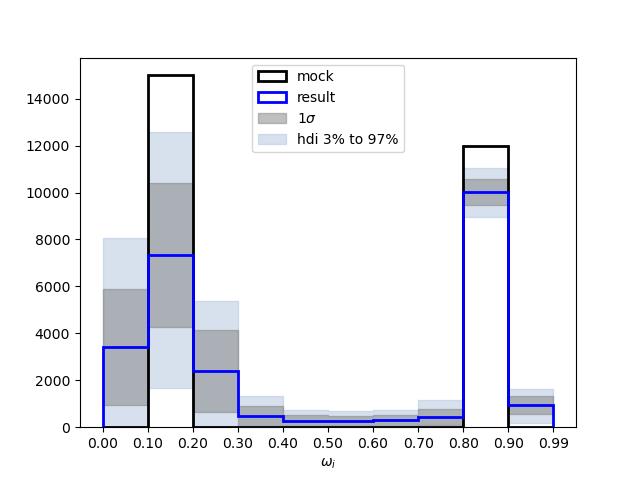}
        \caption{Mock observations made with only one slowly and one fast rotating partial models of NGC 1866.}
        \label{fig:mock_1866_two}
    \end{subfigure}
    \caption{Fits of mock observations computed using the partial models of NGC 419 (left column) and NGC 1866 (right column).}
    \label{fig:mocks_1}
\end{figure*}

\begin{figure*}
    \begin{subfigure}[t]{0.45\linewidth}
        \centering
        \includegraphics[width=\linewidth]{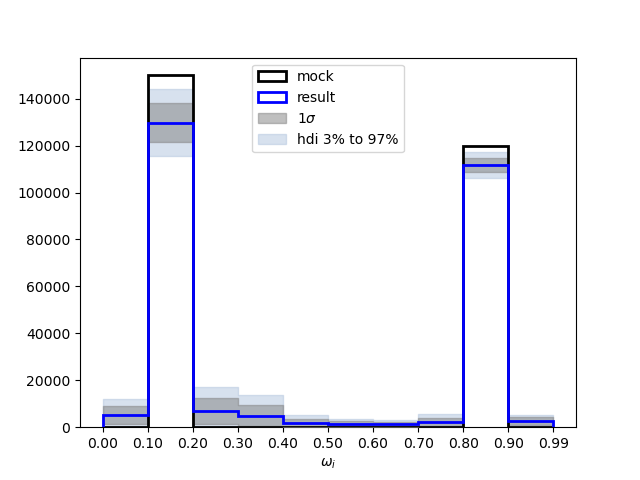}
        \caption{As Figure~\ref{fig:mock_419_two}, with 10 times more stars.}
        \label{fig:mock_419_two_morestars}
    \end{subfigure}\qquad
    \begin{subfigure}[t]{0.45\linewidth}
        \centering
    \includegraphics[width=\linewidth]{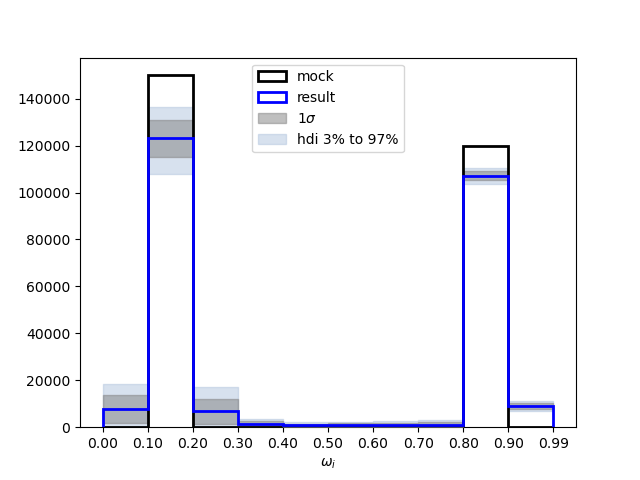}
    \caption{As Figure~\ref{fig:mock_1866_two}, with 10 times more stars.}
    \label{fig:mock_1866_two_morestars}
    \end{subfigure}\
    \begin{subfigure}[t]{0.45\linewidth}
        \centering
        \includegraphics[width=\linewidth]{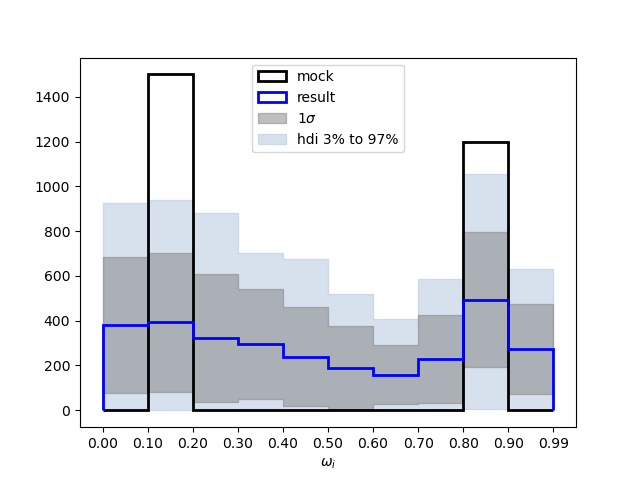}
        \caption{As Figure~\ref{fig:mock_419_two}, with 10 times less stars.}
        \label{fig:mock_419_two_lessstars}
    \end{subfigure}\qquad
    \begin{subfigure}[t]{0.45\linewidth}
        \centering
        \includegraphics[width=\linewidth]{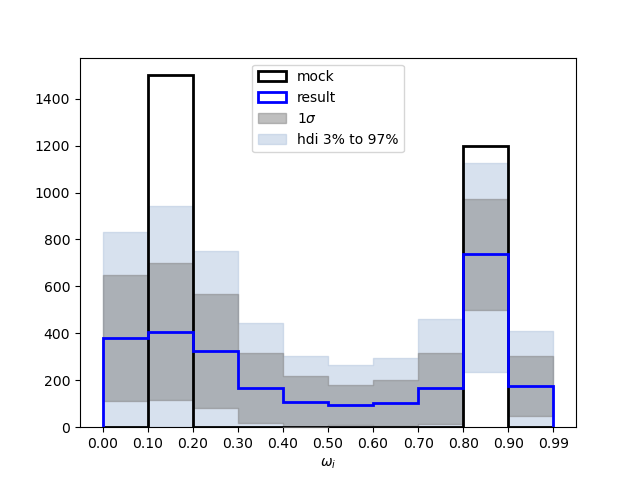}
        \caption{As Figure~\ref{fig:mock_1866_two}, with 10 times less stars.}
        \label{fig:mock_1866_two_lessstars}
    \end{subfigure}\
    \begin{subfigure}[t]{0.45\linewidth}
        \centering
        \includegraphics[width=\linewidth]{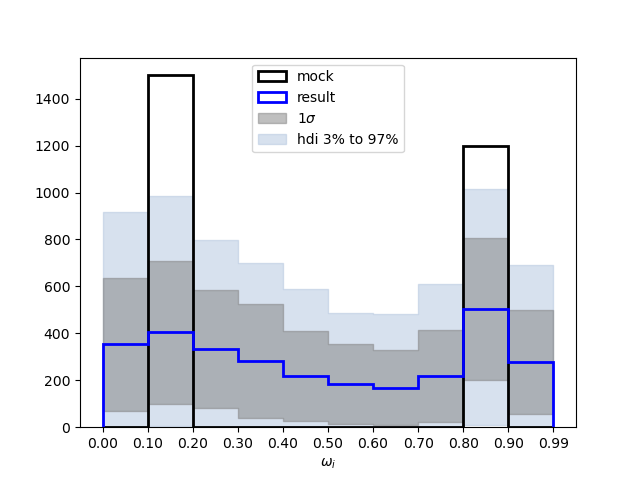}
        \caption{As Figure~\ref{fig:mock_419_two_lessstars}, with double the steps of the Adam optimizer.}
        \label{fig:mock_419_two_lessstars_moresteps}
    \end{subfigure}\qquad
    \begin{subfigure}[t]{0.45\linewidth}
        \centering
        \includegraphics[width=\linewidth]{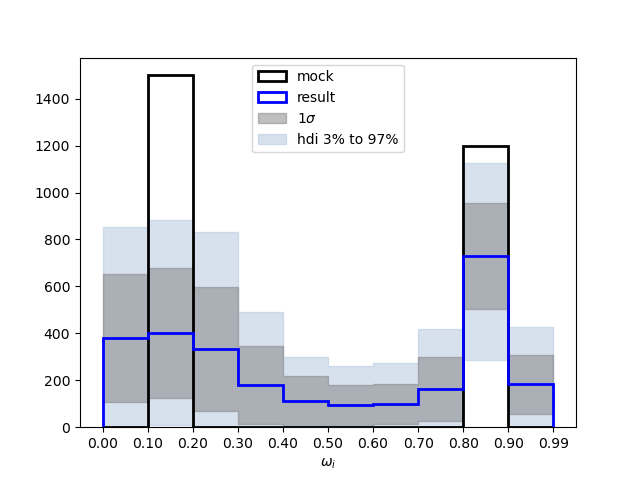}
        \caption{As Figure~\ref{fig:mock_1866_two_lessstars}, with double the steps of the Adam optimizer.}
        \label{fig:mock_1866_two_lessstars_moresteps}
    \end{subfigure}
    \caption{Continuation of Figure~\ref{fig:mocks_1}.}
    \label{fig:mocks_2}
\end{figure*}

\begin{figure*}
    \begin{subfigure}[t]{0.45\linewidth}
        \centering
        \includegraphics[width=\linewidth]{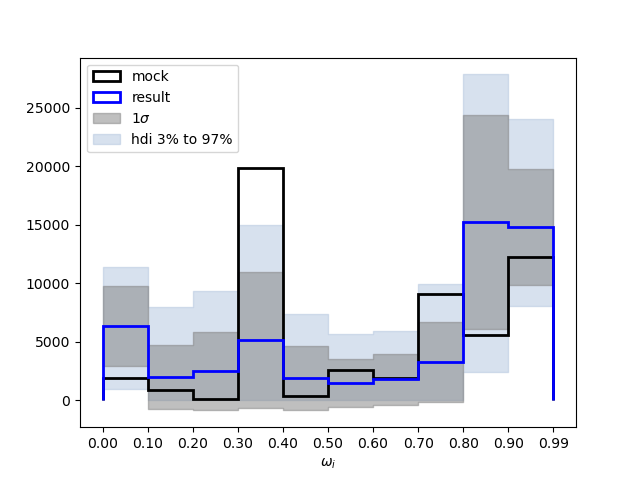}
        \caption{Mock observations made with the partial models of NGC 419 and a 90\% binary fraction.}
        \label{fig:mock_419_fbin90}
    \end{subfigure}\qquad
    \begin{subfigure}[t]{0.45\linewidth}
        \centering
        \includegraphics[width=\linewidth]{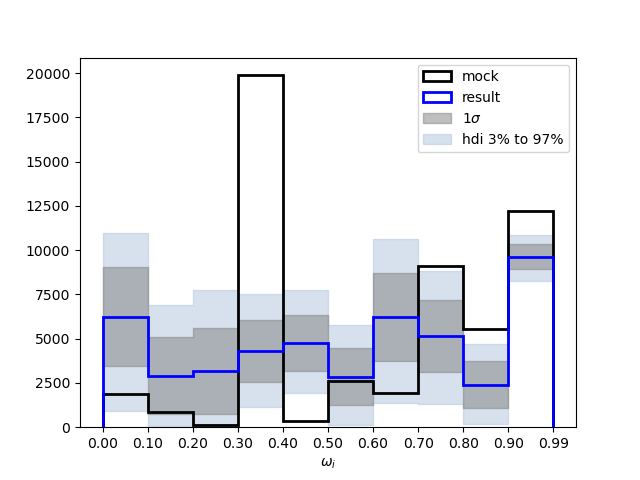}
        \caption{Mock observations made with the partial models of NGC 1866 and a 90\% binary fraction.}
        \label{fig:mock_1866_fbin90}
    \end{subfigure}
    \begin{subfigure}[t]{0.45\linewidth}
        \centering
        \includegraphics[width=\linewidth]{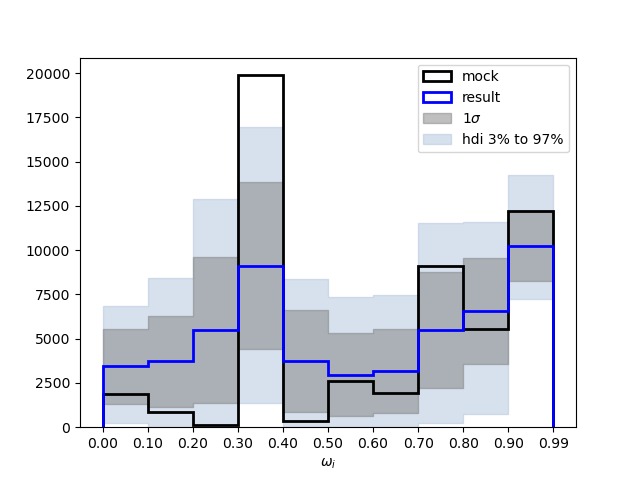}
        \caption{Mock observations made with the partial models of NGC 419 and a 60\% binary fraction.}
        \label{fig:mock_419_fbin60}
    \end{subfigure}\qquad
    \begin{subfigure}[t]{0.45\linewidth}
        \centering
        \includegraphics[width=\linewidth]{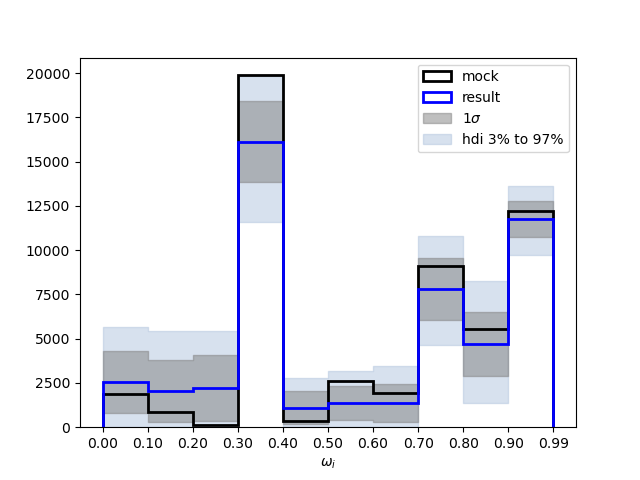}
        \caption{Mock observations made with the partial models of NGC 1866 and a 60\% binary fraction.}
        \label{fig:mock_1866_fbin60}
    \end{subfigure}
    \caption{Continuation of Figure~\ref{fig:mocks_2}.}
    \label{fig:mocks_3}
\end{figure*}

\begin{figure*}
    \centering
    \includegraphics[width=0.45\linewidth]{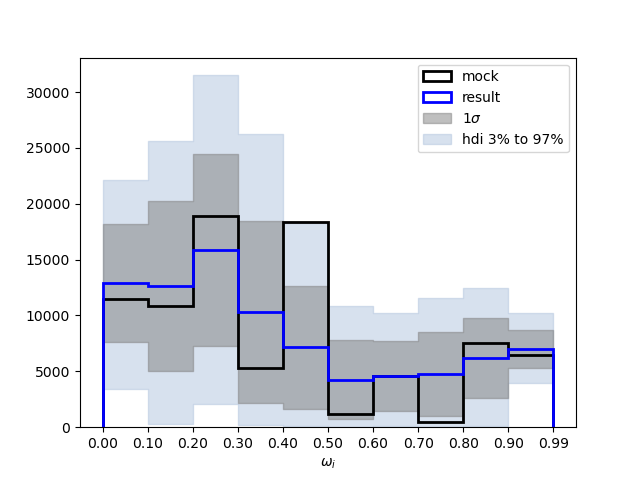}\qquad
    \includegraphics[width=0.45\linewidth]{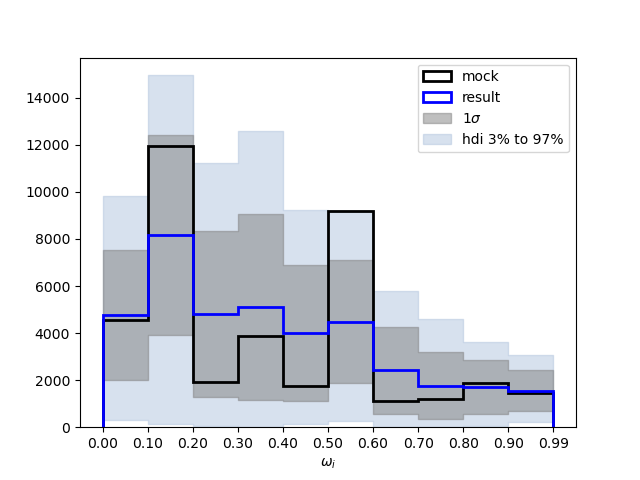}\\
    \includegraphics[width=0.45\linewidth]{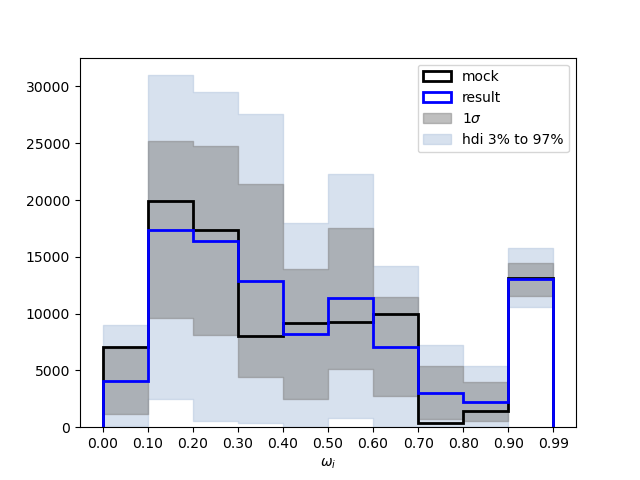}\qquad
    \includegraphics[width=0.45\linewidth]{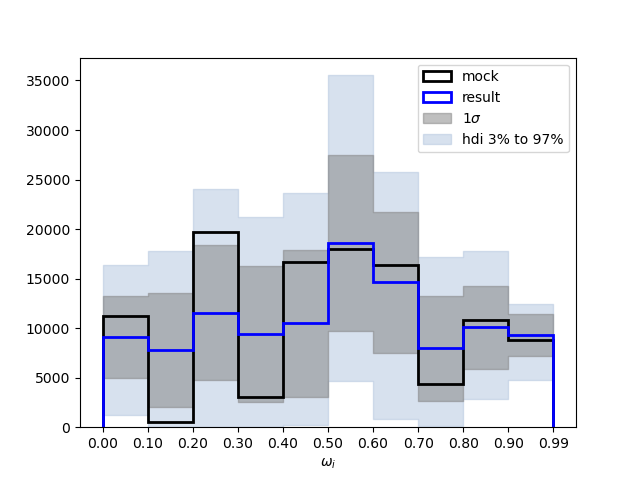}
    \caption{Mocks with partial models for NGC 419 generated with a random velocity distribution.}
    \label{fig:mock_419_random}
\end{figure*}

\begin{figure*}
    \centering
    \includegraphics[width=0.45\linewidth]{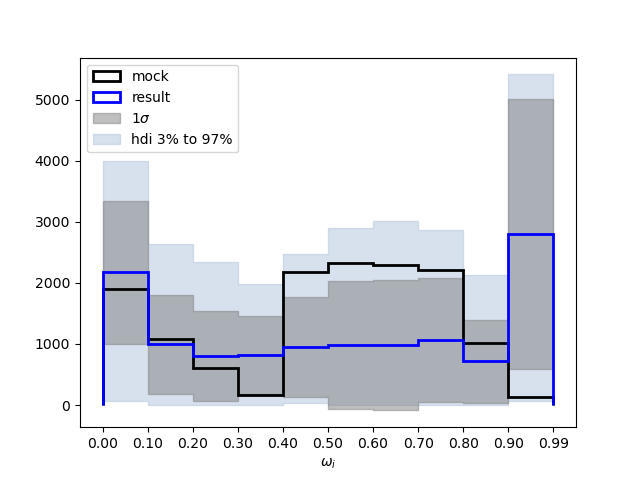}\qquad
    \includegraphics[width=0.45\linewidth]{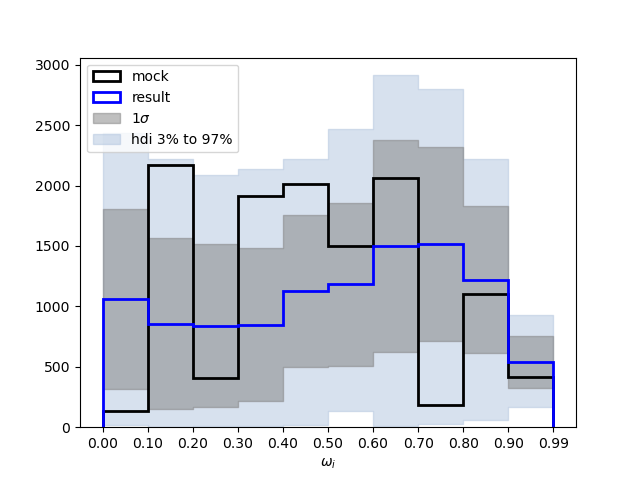}\\
    \includegraphics[width=0.45\linewidth]{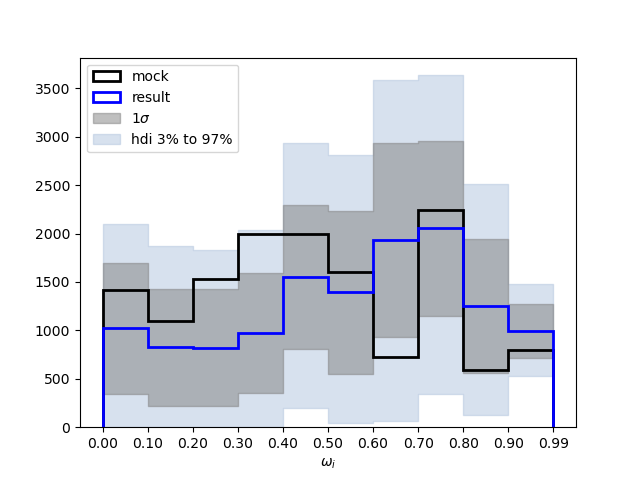}\qquad
    \includegraphics[width=0.45\linewidth]{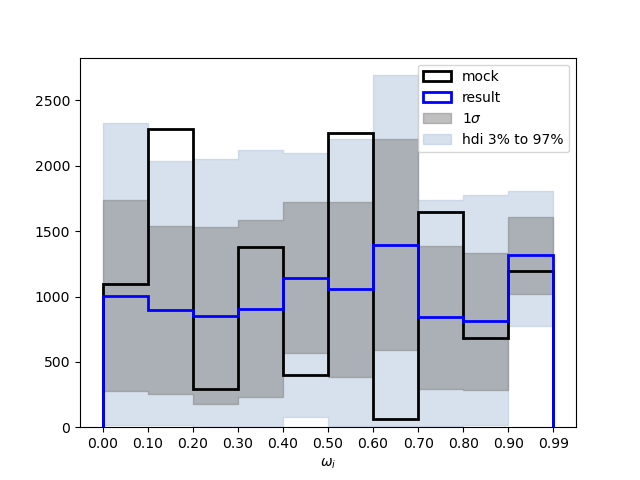}
    \caption{Mocks with partial models for NGC 1866 generated with a random velocity distribution.}
    \label{fig:mock_1866_random}
\end{figure*}
\bsp
\label{lastpage}
\end{document}